\documentclass[12pt]{iopart}
\usepackage{amsthm,amssymb,graphicx}
\usepackage{times}
\usepackage{amsfonts,mathtime}
\newcommand{\Rset}{\mathbb{R}}
\newcommand{\Tset}{\mathbb{T}}
\newcommand{\Zset}{\mathbb{Z}}

\theoremstyle{plain}
\newtheorem{teo}{Theorem}
\newtheorem{pro}[teo]{Proposition}
\newtheorem{lem}[teo]{Lemma}
\newtheorem{cor}[teo]{Corollary}
\newtheorem{defi}{Definition}
\theoremstyle{remark}
\newtheorem{remark}{Remark}
\newtheorem{example}{Example}

\newcommand{\Distance}{\mathop{\rm dist}\nolimits}
\newcommand{\Divergence}{\mathop{\rm div}\nolimits}

\newcommand{\Identity}{{\rm Id}}
\newcommand{\Interior}[1]{i \left( #1 \right)}

\newcommand{\norm}[1]{\left | #1 \right |}
\newcommand{\ce}{{\rm c}}
\newcommand{\st}{{\rm s}}
\newcommand{\un}{{\rm u}}

\begin{document}

\title[Canonical Melnikov theory for diffeomorphisms]
      {Canonical Melnikov theory for diffeomorphisms}

\author{H\'{e}ctor E.~Lomel\'{\i}\dag,
        James D.~Meiss\ddag\ and
        Rafael Ram\'{\i}rez-Ros\S}

\date{today}

\address{\dag  Department of Mathematics,
               Instituto Tecnol\'{o}gico Aut\'{o}nomo de M\'{e}xico,
               Mexico, DF 01000}

\address{\ddag Department of Applied Mathematics,
               University of Colorado,
               Boulder CO 80309-0526}

\address{\S    Departament de Matem\`{a}tica Aplicada I,
               Universitat Polit\`{e}cnica de Catalunya,
               Diagonal 647, 08028 Barcelona, Spain}

\begin{abstract}
We study perturbations of diffeomorphisms that have
a saddle connection between
a pair of normally hyperbolic invariant manifolds.
We develop a first-order deformation calculus for invariant manifolds
and show that a generalized Melnikov function or
\emph{Melnikov displacement} can be written in a canonical way.
This function is defined to be a section of the normal
bundle of the saddle connection.

We show how our definition reproduces the classical methods of
Poincar\'{e} and Melnikov and specializes to methods previously
used for exact symplectic and volume-preserving maps.
We use the method to detect the transverse intersection of stable
and unstable manifolds and relate this intersection to the set of
zeros of the Melnikov displacement.
\end{abstract}



\submitto{Nonlinearity}

\eads{\mailto{lomeli@itam.mx},
      \mailto{jdm@colorado.edu},
      \mailto{rafael@vilma.upc.es}}

\section{Introduction}

The study of the intersections of stable and unstable manifolds of
maps and flows has a strong influence on dynamical systems. In
particular, the existence of a transverse intersection is
associated with the onset of chaos, and gave rise to the famous
horseshoe construction of Smale. The Poincar\'{e}-Melnikov
method~\cite{Poincare99,Melnikov63,HolmesM82} is a widely used
technique for detecting such intersections.  Given a system with a
pair of saddles and a degenerate heteroclinic or saddle connection
between them, the classical Melnikov function computes the rate at
which the distance between the manifolds changes with a
perturbation.

There have been many formulations of the Melnikov method for
two-dimensional maps or flows~\cite{Easton84, GlasserPB89,
DelshamsR96, Lomeli97} and for higher-dimensional symplectic
mappings~\cite{BountisGK95, DelshamsR97,BaldomaF98}. Recently, the
geometric content of Melnikov's method was exploited in order to
detect heteroclinic intersections of Lagrangian manifolds for the
case of perturbed Hamiltonian flows~\cite{Roy06}. Here it was
shown that the heteroclinic orbits are in correspondence with the
zeros of a geometric object, the so-called Melnikov one-form.

For maps, the Melnikov function is an infinite sum whose domain is
a saddle connection between two hyperbolic invariant sets. As
usual, a simple zero of this function corresponds to a transverse
intersection of stable and unstable manifolds of a perturbation of
the original map.

Melnikov's method can also be used to compute transport fluxes. In
particular, a \emph{resonance zone} for a two-dimensional mapping
is a region bounded by alternating segments of stable and unstable
manifolds that are joined at primary intersection
points~\cite{MacKayMP87,Easton91}. Because the intersection points
are primary, a resonance zone is bounded by a Jordan curve and has
exit and entry sets~\cite{EastonMC93}. The images of these sets
completely define the transport properties of the resonance zone.
Moreover, the integral of the Melnikov function between two
neighbouring primary intersection points is the first order
approximation to the geometric flux escaping from the resonance
zone~\cite{MacKayM88, KaperW91}.

The method has also been applied to the case of periodically
time-dependent, volume-preserving flows~\cite{MezicW94} and more
generally to volume-preserving maps with fixed
points~\cite{LomeliM00a} and invariant circles~\cite{LomeliM03}.
Volume-preserving maps provide perhaps the simplest, natural
generalization of the class of area-preserving maps to higher
dimensions. Moreover, they naturally arise in applications as the
time-one Poincar\'{e} map of incompressible flows---even when the
vector field of the flow is nonautonomous. Thus the study of the
dynamics of volume-preserving maps has application both to fluids
and magnetic fields.

Our goal in this paper is to develop, based on the theory of
deformations, a general, geometrical description of the Melnikov
displacement and to compare our theory to classical results.
Deformation theory was first introduced in the theory of
singularities~\cite{ThomL71}, but was soon used in the contexts of
volume and symplectic geometry. Its application to dynamical
systems in~\cite{LlaveMM86,Llave96,DelshamsLS06} provide results
that are close to our goals.

Let $f_\epsilon$ be a smooth family of diffeomorphisms such that
the unperturbed map $f_0$ has a saddle connection $\Sigma$ between
a pair of compact $r$-normally hyperbolic invariant manifolds. Let
$\nu(\Sigma) \equiv T_\Sigma M/T \Sigma$ be the algebraic normal
bundle of the saddle connection. We show that there exists a
canonical $C^{r-1}$ section $\mathcal{D}:\Sigma \to \nu(\Sigma)$,
called the \emph{Melnikov displacement}, that measures the
splitting of the saddle connection in first-order. We will prove
that the Melnikov displacement is given by the absolutely
convergent series
\begin{equation}\label{mainformula}
\mathcal{D} =
\sum_{k\in\Zset}(f_0^\ast)^k \mathcal{F}_0 =
\sum_{k\in\Zset}({f_0}_\ast)^k \mathcal{F}_0
\end{equation}
where $\mathcal{F}_\epsilon$ is the vector field defined by
$\frac{\partial}{\partial \epsilon} f_\epsilon =
\mathcal{F}_\epsilon \circ f_\epsilon$.

These sums do not converge in the tangent space $T_\Sigma M$, but
only in the algebraic normal bundle $\nu(\Sigma)$. The use of the
algebraic normal bundle in the study of normally hyperbolic
manifolds goes back to~\cite{HirschPS77}. The fact that
(\ref{mainformula}) always converges also addresses the question
of how to deal with Melnikov's method when the original pair of
compact $r$-normally hyperbolic invariant manifolds (that are
connected with the saddle connection) are not fixed with the
perturbation.

In addition, we will also show that the Melnikov displacement has
a number of geometric properties. The main result in this
direction is that any change of coordinates acts on the
displacement by its pullback. This result will be used to obtain
the natural action of any symmetries, reversing symmetries or
integrals of the dynamical system on the displacement. Similarly,
if the map preserves a symplectic or volume form, this gives
additional structure to the displacement. For example, if
$f_\epsilon$ is a family of exact symplectic maps and the normally
hyperbolic invariant manifolds are fixed points, then we will show
that there exists a function $L:\Sigma \to \Rset$, the
\emph{Melnikov potential}, such that ${\rm d} L =
\Interior{\mathcal{D}} \omega$, where $\omega$ is the symplectic
two-form. This relation is reminiscent of the definition of
globally Hamiltonian vector fields. When the normally hyperbolic
invariant manifolds are not fixed points (or isolated periodic
points), its stable and unstable manifolds are coisotropic, but
not isotropic, and so the relation ${\rm d} L =
\Interior{\mathcal{D}} \omega$ makes no sense.

We complete this introduction with a note on the organization of
this paper. The general theory is developed in
\S\ref{sec:MelnikovDisplacement}. In \S\ref{sec:classical}, we
show how our theory reproduces the classical methods of Poincar\'e
and Melnikov. The study of exact symplectic maps and
volume-preserving maps is contained in \S\ref{sec:symp} and
\S\ref{sec:volume}, respectively.

\section{Melnikov displacement}\label{sec:MelnikovDisplacement}

\subsection{Deformation calculus}

In this subsection we present the deformation calculus for
families of diffeomorphisms and submanifolds. We shall begin with
diffeomorphisms by defining a vector field associated with a
deformation. Next, we construct a vector field for the deformation
of (immersed) submanifolds. Finally, we will combine these results
to define and compute the Melnikov displacement.

In this paper,
we consider smooth families of diffeomorphisms $f_\epsilon:M \to M$,
where $M$ is an $n$-dimensional smooth manifold.
Here, the term smooth family means that
$f_\epsilon(\xi) \equiv f(\xi,\epsilon)$ is $C^\infty$ in both variables.
The map $f_0$ will be called ``unperturbed".

\begin{defi}[Generating Vector Field~\cite{LlaveMM86}]
The generating vector field of a smooth family of diffeomorphisms
$f_\epsilon$ is the unique vector field $\mathcal{F}_\epsilon$ such that
\begin{equation}\label{eq:pertvect}
\frac{\partial}{\partial\epsilon} f_\epsilon = \mathcal{F}_\epsilon \circ f_\epsilon \;.
\end{equation}
\end{defi}

If we regard $\mathcal{F}_\epsilon$ as a nonautonomous vector field with time $\epsilon$,
then the function $\Phi_{t,s}=f_t \circ f_s^{-1}$ represents its
flow~\cite[Thm.~2.2.23]{AbrahamM78}.
Consequently,
if $f_\epsilon$ is volume preserving,
then $\mathcal{F}_\epsilon$ has zero divergence~\cite[Thm.~2.2.24]{AbrahamM78},
and if $f_\epsilon$ is (exact) symplectic,
then $\mathcal{F}_\epsilon$ is (globally) Hamiltonian.
These geometric equivalences form the basis of the deformation calculus.

\begin{remark}
The generating vector field $\mathcal{F}_\epsilon$ was also called
the {\em perturbation vector field} in~\cite{LomeliM00a};
however, we adopt the older terminology here.
Sometimes,
we will also refer to $\mathcal{F}_\epsilon$ as the \emph{generator}
of the family $f_\epsilon$.
\end{remark}

We will always use the convention that,
given a family of diffeomorphisms denoted by italic letters $f_\epsilon$,
its generator is denoted by the same letter in calligraphic capitals.
We have collected in the proposition below some well-known relations
among generators, in which appear push-forwards and pullbacks.
Recall that the pullback $f^\ast$ and push-forward $f_\ast$ of
a diffeomorphism $f:M\to M$ act on a vector field $\mathcal{X}:M \to TM$
as follows:
\begin{eqnarray*}
f^\ast \mathcal{X} & = &
{\rm D} f^{-1} \circ f  \mathcal{X} \circ f =
({\rm D} f^{-1} \mathcal{X})\circ f \\
f_\ast \mathcal{X} & = &
{\rm D} f \circ f^{-1}  \mathcal{X} \circ f^{-1} =
({\rm D} f \mathcal{X} ) \circ f^{-1} \; .
\end{eqnarray*}
We note that $f_\ast= (f^\ast)^{-1} = (f^{-1})^\ast$.

Next,
our goal is to develop a first-order deformation calculus for invariant
manifolds of smooth diffeomorphisms.
We are mainly interested in stable and unstable invariant manifolds of
$r$-normally hyperbolic manifolds for some $r \ge 1$,
which unfortunately are just
immersed submanifolds (not embedded submanifolds), and $C^r$ (not $C^\infty$).
This gives rise to a few technicalities. Recall that a map $g:N \to M$ is
an {\em immersion} when its differential has maximal rank everywhere.
If $g$ is one-to-one onto its image,
then $W=g(N)$ is an immersed submanifold of the same dimension as $N$.
We will denote an immersion by $W=g(N)\hookrightarrow M$
or simply $W\hookrightarrow M$ when the immersion $g$ does not matter.
For brevity, we will sometimes omit the term ``immersed".

\begin{example}
If $W$ is the stable (resp., unstable) invariant manifold of a
fixed point of a diffeomorphism, then $N=\Rset^s$ (resp., $N=\Rset^u$),
where $s$ and $u$ are the number of stable and unstable directions at the
hyperbolic point. When the fixed point is hyperbolic, $n=s+u$.
Stable and unstable manifolds are typically not embedded because they can have
points of accumulation. In this case the immersion is not a proper map.
\end{example}

We consider families of submanifolds of the form
$W_\epsilon = g_\epsilon(N) \hookrightarrow M$,
where $g(\xi,\epsilon)$ is $C^r$ in both variables, for some $r\ge 1$.
All the elements of such a family are diffeomorphic (as immersed submanifolds),
because they are diffeomorphic to the same ``base" manifold $N$.
Just as for $f_0$, the unperturbed submanifold is denoted by $W_0$.

\begin{defi}[Adapted Deformation]\label{def:adapted}
If $W_\epsilon \hookrightarrow M$ is a $C^r$ family of immersed
submanifolds,
a family of diffeomorphisms $\phi_\epsilon: W_0 \to W_\epsilon$ is
an adapted deformation when $\phi_0=\Identity_{W_0}$ and
$\phi(\xi,\epsilon)$ is $C^r$ in both variables.
\end{defi}

Adapted deformations exist since
it suffices to take $\phi_\epsilon=g_\epsilon \circ g_0^{-1}$.
While there is quite a bit of freedom in the choice of $\phi_\epsilon$,
only its normal component is relevant,
since this measures the actual motion of $W_\epsilon$ with $\epsilon$.
The normal component will be defined using
the algebraic normal bundle. For an immersed submanifold $W \hookrightarrow M$,
this is defined as the set of equivalence classes
\begin{equation}\label{eq:normalbundle}
\nu(W) \equiv T_{W}M\diagup TW \;.
\end{equation}
When $M$ is Riemannian, this normal bundle is isomorphic to the more familiar
geometric normal bundle, $TM^{\bot}$ (cf.~\cite[p.96]{Hirsch76}).
In general $\nu$ is a manifold of dimension $n$ and is defined independently
of any inner product structure on $TM$.

\begin{defi}[Displacement Vector Field]
The displacement vector field of a $C^r$ family of immersed submanifolds
$W_\epsilon \hookrightarrow M$ is the $C^{r-1}$ section
\begin{equation}\label{eq:disp}
\mathcal{D}:W_0\rightarrow\nu(W_0) \;, \qquad
\mathcal{D}(\xi) \equiv
\left.\frac{\partial}{\partial\epsilon}\right|_{\epsilon=0}
\phi_\epsilon(\xi) + T_{\xi}W_0
\end{equation}
where $\phi_\epsilon:W_0 \to W_\epsilon$ is any adapted deformation.
\end{defi}

The displacement vector field is well-defined;
that is, its definition is independent of the choice of the adapted deformation
as is shown in the following lemma.

\begin{lem}\label{lem:adapted}
Let $\phi_\epsilon,\tilde{\phi}_\epsilon:W_0 \to W_\epsilon$ be
two adapted deformations. Then
\[
\left[
\frac{\partial}{\partial\epsilon} \tilde\phi_\epsilon(\xi) -
\frac{\partial}{\partial\epsilon} \phi_\epsilon(\xi)
\right]_{\epsilon=0} \in T_\xi W_0 \;,\qquad
\forall \xi \in W_0 \;.
\]
\end{lem}

\proof
For each fixed $\xi \in W_0$, the map
$\epsilon \mapsto c(\epsilon)\equiv\phi_\epsilon^{-1}(\tilde{\phi}_\epsilon(\xi))$
describes a $C^r$ curve in $W_0$ such that
$c(0)=\xi$ and $c'(0) \in T_\xi W_0$.
Using  the fact that $\tilde{\phi}_\epsilon(\xi)=\phi_\epsilon(c(\epsilon))$,
we then have
\[
\left[
\frac{\partial}{\partial\epsilon} \tilde\phi_\epsilon(\xi) -
\frac{\partial}{\partial\epsilon} \phi_\epsilon(\xi)
\right]_{\epsilon=0} =
{\rm D} \phi_0(\xi) c'(0) = c'(0) \in T_\xi W_0 \;,
\]
because ${\rm D} \phi_0(\xi)=\Identity_{T_\xi W_0}$.
\qed

When the submanifold is invariant under a diffeomorphism,
its deformations are related by means of a fundamental iterative
relationship between the generating vector field of the family of
diffeomorphisms and the displacement vector field of the family of
submanifolds.

\begin{pro}
Let $f_\epsilon$ be a smooth family of diffeomorphisms, and
$W_\epsilon \hookrightarrow M$ be a $C^r$ family of immersed submanifolds
that are invariant under $f_\epsilon$.
Then
\begin{equation}\label{eq:iterate}
f_0^\ast \mathcal{D} - \mathcal{D} = f_0^\ast \mathcal{F}_0
\end{equation}
on the unperturbed submanifold $W_0$,
where $\mathcal{D}$ is the displacement vector field~\eref{eq:disp}.
\end{pro}

\proof
The tangent space $T W_0$ is invariant under the pullback $f_0^\ast$,
so the term $f_0^\ast\mathcal{D}$ is well-defined as a section of the normal
bundle $\nu(W_0)$.
If $\phi_\epsilon: W_0 \to W_\epsilon$ is any adapted deformation, then
$\tilde{\phi}_\epsilon \equiv f_\epsilon \circ\phi_\epsilon\circ f_0^{-1}$ is
as well. Differentiating
$\tilde{\phi}_\epsilon \circ f_0 =
 f_\epsilon \circ \phi_\epsilon$ with respect to $\epsilon$ yields
\[
\left.  \frac{\partial}{\partial\epsilon}\right|_{\epsilon=0}
\tilde{\phi}_\epsilon \circ f_0 =
\mathcal{F}_0 \circ f_0 + {\rm D} f_0 \left.
\frac{\partial}{\partial\epsilon}\right|_{\epsilon=0} \phi_\epsilon
\]
where we used~\eref{eq:pertvect}.
However, by lemma~\ref{lem:adapted} the displacement~\eref{eq:disp}
is independent of the adapted deformation, so
\[
\mathcal{D} \circ f_0 =
\mathcal{F}_0 \circ f_0 + {\rm D} f_0 \mathcal{D} \;.
\]
Applying ${\rm D} f_0^{-1} \circ f_0$ to both sides finishes the proof.
\qed

The identity~\eref{eq:iterate} is equivalent to $\mathcal{D}=
(f_0)_\ast \mathcal{D} + \mathcal{F}_0$. Thus, we can work either
with push-forwards or pullbacks. To obtain the Melnikov
displacement we will iterate these identities on the stable and
unstable manifolds of a family of diffeomorphisms.

\subsection{Normally hyperbolic invariant manifolds and saddle connections}

The Melnikov displacement will be defined for a saddle connection
between a pair of normally hyperbolic invariant manifolds. In
this section we recall the definitions of these objects.
There are many slightly different definitions of normally
hyperbolic manifolds, see~\cite{HirschPS77}.
In this paper, we adopt the following.

\begin{defi}[Normally Hyperbolic Invariant Manifold]
Let $A \subset M$ be a submanifold invariant under a smooth
diffeomorphism $f:M \to M$ .
We say that $A$ is $r$-normally hyperbolic
when there exist a Riemann structure on $TM$,
a constant $\lambda \in (0,1)$,
and a continuous  invariant splitting
\begin{equation}\label{eq:splitting}
T_a M = E_a^\st \oplus E_a^\un \oplus T_a A \; ,
\qquad \forall a \in A
\end{equation}
such that if $L^{\st,\un}_a : E^{\st,\un}_a \to E^{\st,\un}_{f(a)}$
and $L^\ce_a : T_a A \to T_{f(a)} A$ are the
canonical restrictions of the linear map
${\rm D}f(a): T_a M \to T_{f(a)} M$
associated to the splitting~\eref{eq:splitting},
then
\begin{enumerate}
\item
$\norm{(L^\ce_a)^{-1}}^l \norm{ L^\st_a } <  \lambda$, and
\item
$\norm { L^\ce_a}^l \norm{(L^\un_a)^{-1}}< \lambda$
\end{enumerate}
for all $l=0,1,\ldots,r$, and for all $a \in A$.
\end{defi}

As is usual in the literature, the term \emph{normally hyperbolic}
will be taken to mean $1$-\emph{normally hyperbolic}. Note that
setting $l=0$ in the previous definition implies that the
linearizations $L^\st$ and $L^\un$ of $f$ restricted to the stable
and unstable spaces of $A$ have the uniform bounds
\begin{equation}\label{eq:LuLs}
\norm{L^\st_a},\norm{(L^\un_a)^{-1}} < \lambda < 1\; ,\qquad
\forall a\in A\;.
\end{equation}

In this paper
we will assume that each $r$-normally hyperbolic invariant manifold $A$ is compact,
although it would be sufficient that our diffeomorphisms be
uniformly $C^r$ in some neighbourhood of $A$.
We will also assume, without loss of generality, that $A$ is connected.
One consequence is that the sets
\begin{equation*}
W^\st = W^\st(A) = W^\st(A,f)  =  \left\{ \xi \in M : \lim_{k\to
+\infty}\Distance(f^k(\xi),A)=0 \right\}
\end{equation*}
and
\begin{equation*}
W^\un = W^\un(A) = W^\un(A,f)  =  \left\{ \xi \in M : \lim_{k\to
-\infty}\Distance(f^k(\xi),A)=0 \right\}
\end{equation*}
are $C^r$ immersed submanifolds of $M$ that are tangent at $A$
to  $T A \oplus E^{\st,\un}$, see~\cite{HirschPS77}.
In particular, $T_A W^\st \cap T_A W^\un = T A$.
Moreover, $A$ and its
stable and unstable invariant manifolds are persistent:
given any smooth family of diffeomorphisms $f_\epsilon$ such that $f=f_0$,
then for each small enough $\epsilon$ there exists a nearby $r$-normally
hyperbolic invariant manifold $A_\epsilon$
with $C^r$ families of immersed submanifolds
$W^{\st,\un}_\epsilon=W^{\st,\un}(A_\epsilon,f_\epsilon)$.

To compute the Melnikov displacement, we will need
to show that certain series are geometrically convergent;
the following lemma is a key component in this proof.

\begin{lem}\label{lem:LNBound}
Let $f: M \to M$ be a diffeomorphism with a compact normally hyperbolic
invariant manifold $A$, and fix a point $\xi \in W^\st = W^\st(A,f)$.
Then given any splitting $T_\xi M = T_\xi W^\st \oplus N_\xi$,
there exists a constant $\mu \in (0,1)$ and an integer $n_0 > 0$
such that
\[
\norm{(L^N_n)^{-1}} < \mu , \; \qquad \forall n \ge n_0
\]
where $L^N_n : N_{\xi_n} \to N_{\xi_{n+1}}$ are the restrictions
of ${\rm D}f(\xi_n): T_{\xi_n} M \to T_{\xi_{n+1}} M$ to the
subspaces $N_{\xi_n} = {\rm D}f^n(\xi)[N_\xi]$, $\xi_n = f^n(\xi)$.
A similar bound holds for $W^\un$.
\end{lem}
\proof
There exists a unique point $a \in A$ such that
$\lim_{n\to +\infty}\Distance(\xi_n,a_n)=0$,
where $a_n=f_0^n(a)$.
According to the $\lambda$-Lemma for normally hyperbolic manifolds~\cite{CressonW05},
the complementary subspaces $N_{\xi_n}$ tend to
the unstable subspaces $E^\un_{a_n}$, as $n\to +\infty$.
Thus, the maps $L_n^N : N_{\xi_n} \to N_{\xi_{n+1}}$
tend to the unstable restrictions
$L^\un_{a_n}: E^\un_{a_n} \to E^{\un}_{a_{n+1}}$ as $n\to +\infty$,
and the lemma follows from~\eref{eq:LuLs}.
It suffices to take any $\mu \in (\lambda,1)$.
\qed

The Melnikov displacement will be defined
as a function on the normal
bundle of a \emph{saddle connection},
which is defined as follows.

\begin{defi}[Saddle Connection]
Let $f: M \to M$ be a diffeomorphism with a pair of
compact normally hyperbolic invariant manifolds $A$ and $B$.
A saddle connection between $A$ and $B$ is an invariant submanifold
$\Sigma \subset W^\un(A) \cap W^\st(B)$ such that
\[
T_\xi \Sigma = T_\xi W^\un(A) = T_\xi W^\st(B)
\]
for all $\xi \in \Sigma$.
\end{defi}

\begin{remark}\label{rem:normalbundles}
The coincidence of the tangent spaces is needed in order that the
manifolds have the same algebraic normal bundles on the saddle
connection:
\[
\nu(\Sigma) =
\nu(W^\un(A))|_\Sigma =
\nu(W^\st(B))|_\Sigma
\]
since
$T_\xi M / T_\xi \Sigma =
 T_\xi M / T_\xi W^\un(A) =
 T_\xi M / T_\xi W^\st(B)$,
for all $\xi \in \Sigma$.
\end{remark}

By definition, $\dim \Sigma = \dim W^\un(A) = \dim W^\st(B)$, and
the manifolds $A$ and $B$ are not part of the saddle connection.
The simplest (and most common) saddle connections are of the form
$\Sigma = W^\un(A) \setminus A = W^\st(B) \setminus B$. In this
case, we say that the unperturbed invariant manifolds are
\emph{completely doubled}. Many Melnikov problems studied in the
literature fall into this category. Nevertheless, in some problems
there may exist points $\xi \in  W^\un(A) \cap W^\st(B)$, $\xi
\not \in A \cup B$, such that $T_\xi  W^\un(A) \neq T_\xi
W^\st(B)$, see~\cite{DelshamsR97}. In that case the saddle
connection is strictly contained in the intersection of the stable
and unstable invariant manifolds: $\Sigma \varsubsetneq (W^\un(A)
\cap W^\st(B)) \setminus (A \cup B)$.

\subsection{Displacement vector fields of stable and unstable
            invariant manifolds}

In this subsection we use the fundamental iterative
equation~\eref{eq:iterate} to obtain infinite series
for the displacements of the stable and unstable manifolds.
These series are absolutely convergent, but only,
as we must stress, when they are evaluated on their corresponding
normal bundles.
Indeed, the tangential components of these series can be unbounded.
Consequently, in order to compute these sums,
each term must be projected onto the normal bundle.
An example will be given in \S\ref{ssec:Suris}.

The proof of the following proposition is inspired by a proof
given in~\cite{BaldomaF98},
the main difference is that our setting is more geometric.

\begin{pro}\label{pro:DuDs}
Let $f_\epsilon$ be a smooth family of diffeomorphisms such that the
unperturbed map $f_0$ has a compact normally hyperbolic invariant
manifold $A_0$ with stable and unstable invariant manifolds $W^{\st,\un}_0$.
Then the displacement vector fields
$\mathcal{D}^\st:W^\st_0 \to \nu(W^\st_0)$ and
$\mathcal{D}^\un:W^\un_0 \to \nu(W^\un_0)$
of the families of perturbed stable and unstable invariant manifolds
are given by the absolutely convergent series
\[
\mathcal{D}^\st = -\sum_{k \ge 1} (f_0^\ast)^k \mathcal{F}_0 \;, \qquad
\mathcal{D}^\un = \sum_{k \le 0} (f_0^\ast)^k \mathcal{F}_0 \;.
\]
\end{pro}

\proof
We prove the claim about the stable displacement $\mathcal{D}^\st$;
the unstable result is obtained analogously.
Repeatedly applying the iterative formula~\eref{eq:iterate}, yields
\[
\mathcal{D}^\st =
(f^\ast_0)^n \mathcal{D}^\st - \sum_{k=1}^n (f_0^\ast)^k \mathcal{F}_0
\]
for any integer $n \ge 1$.
Therefore,
it suffices to check that the term $(f^\ast_0)^n \mathcal{D}^\st$
tends geometrically to zero on the normal bundle of the unperturbed stable
manifold. For any point $\xi \in W^\st_0 = W^\st(A_0,f_0)$, let
$\xi_n \equiv f_0^n(\xi)$, and
$\mathcal{D}^\st_n \equiv \mathcal{D}^\st(\xi_n) \in T_{\xi_n} M$. Then
$(f^\ast_0)^n \mathcal{D}^\st(\xi) = (\bar{L}_n)^{-1} \mathcal{D}^\st_n$
where $\bar{L}_n \equiv {\rm D} f_0^n(\xi) : T_\xi M \to T_{\xi_n} M$.
We must show that $(\bar{L}_n)^{-1} \mathcal{D}^\st_n $ tends geometrically
to zero as an element of
the quotient space $T_\xi M/T_\xi W^\st_0$.

Given any splitting $T_\xi M = T_\xi W^\st_0 \oplus N_\xi$,
let $\Pi^N_n: T_{\xi_n} M \to N_{\xi_n}$ be the projections
onto the linear subspaces $N_{\xi_n} = {\rm D}f_0^n(\xi)[N_\xi]$,
and let $L^N_n : N_{\xi_n} \to N_{\xi_{n+1}}$ be the corresponding restrictions
of the linear maps
$L_n \equiv {\rm D}f_0(\xi_n): T_{\xi_n} M \to T_{\xi_{n+1}} M$.
Then,
\begin{eqnarray*}
\Pi^N_0 (\bar{L}_n)^{-1} \mathcal{D}^\st_n
& = &
\Pi^N_0 (L_0)^{-1} (L_1)^{-1} \cdots (L_{n-1})^{-1} \mathcal{D}^\st_n \\
& = &
(L^N_0)^{-1} (L^N_1)^{-1} \cdots (L^N_{n-1})^{-1} \Pi^N_n \mathcal{D}^\st_n
\end{eqnarray*}
because $\bar{L}_n = L_{n-1} \cdots L_1 L_0$ and
$L^N_k \circ \Pi^N_k = \Pi^N_{k+1} \circ L^N_k$.
Let $n_0$ be the integer referred
to in lemma~\ref{lem:LNBound}, and define  $l_{n_0}=\Pi_{k=0}^{n_0-1} \norm{(L^N_k)^{-1}}$.
Lemma~\ref{lem:LNBound} gives the bound
\[
\norm{\Pi^N_0 (\bar{L}_n)^{-1} \mathcal{D}^\st_n} \le
l_{n_0} \mu^{n-n_0} \norm{\Pi^N_n \mathcal{D}^\st_n} \le
l_{n_0} \mu^{n-n_0} \norm{\mathcal{D}^\st_n}
\]
for some $\mu \in (0,1)$.
Moreover,
the sequence $(\mathcal{D}^\st_n)_{n\ge 0}$ is bounded due to
the compactness of the normally hyperbolic manifold $A_0$
and the continuity of the displacement vector field
$\mathcal{D}^\st(\xi)$.
Therefore, $\norm{\Pi^N_0 (\bar{L}_n)^{-1} \mathcal{D}^\st_n}$
tends geometrically to zero as $n\to 0$,
and so,
$(\bar{L}_n)^{-1} \mathcal{D}^\st_n $ tends geometrically to zero in
the quotient space $T_\xi M/T_\xi W^\st_0$.
\qed

\subsection{Melnikov displacement}\label{sec:meldis}

We now will use the displacement
vector fields to study the splitting of a saddle connection upon perturbation.
As usual, $f_\epsilon$ denotes a smooth family of diffeomorphisms such that the
unperturbed map $f_0$ has a saddle connection
$\Sigma \subset W^{\un}(A_0,f_0) \cap W^{\st}(B_0,f_0)$ between a pair of
compact $r$-normally hyperbolic invariant manifolds $A_0$ and $B_0$.
These manifolds persist and remain $r$-normally hyperbolic for small $\epsilon$.

We want to study the distance between the perturbed manifolds
$W^{\un}_\epsilon = W^{\un}(A_\epsilon,f_\epsilon)$
and $W^{\st}_\epsilon = W^{\st}(B_\epsilon,f_\epsilon)$
The growth rate of this distance with $\epsilon$ is obtained
simply by taking the difference between the displacement vector fields of
both families.

\begin{defi}[Melnikov Displacement]\label{def:melnikov}
Under the previous assumptions, the Melnikov displacement is the
canonical $C^{r-1}$ section of the normal bundle $\nu(\Sigma)$
defined by
\begin{equation}\label{eq:melnikovDisp1}
\mathcal{D} \equiv \mathcal{D}^\un-\mathcal{D}^\st :\Sigma\rightarrow\nu(\Sigma)
\end{equation}
where $\mathcal{D}^\un:W^\un_0 \to \nu(W^\un_0)$ and
$\mathcal{D}^\st:W^\st_0 \to \nu(W^\st_0)$ are the displacement vector fields of
$W^\un_\epsilon$ and $W^\st_\epsilon$, respectively.
\end{defi}

\begin{remark}
The Melnikov displacement $\mathcal{D}$ makes sense only on the saddle
connection,  where the tangent spaces of the unperturbed invariant manifolds
$W^\un_0$ and $W^\st_0$ coincide.
Away from $\Sigma$, the difference $\mathcal{D}^\un - \mathcal{D}^\st$ is undefined
because each term is a section of a different (algebraic) normal bundle,
see remark~\ref{rem:normalbundles}.
\end{remark}

\begin{cor}
Let $f_\epsilon:M \to M$ be a smooth family of diffeomorphisms verifying the
assumptions of definition~\ref{def:melnikov}.
Then its Melnikov displacement $\mathcal{D}: \Sigma \to \nu(\Sigma)$
is given by the absolutely convergent sums
\begin{equation}\label{eq:MelnikovSums}
\mathcal{D} =
\sum_{k\in\Zset} (f_0^\ast)^k \mathcal{F}_0 =
\sum_{k\in\Zset} ({f_0}_\ast)^k \mathcal{F}_0 \;.
\end{equation}
\end{cor}

\proof
 From proposition~\ref{pro:DuDs}, we get that
$\mathcal{D} = \mathcal{D}^\un - \mathcal{D}^\st =
 \sum_{k\in\Zset} (f_0^\ast)^k \mathcal{F}_0 =
 \sum_{k\in\Zset} ({f_0}_\ast)^k \mathcal{F}_0$,
where the last equality follows from the identity
$({f_0}_\ast)^k=(f_0^\ast)^{-k}$.
\qed

The Melnikov displacement has been defined in a canonical way as a
section of the normal bundle,
as such it has strong geometric properties.
We will show next that any change of variables acts as a pullback on it.
This will imply a number of geometrical properties,
for example that the displacement is invariant under the pullback and
the push-forward of the unperturbed map.
In addition, we will see that if there exist symmetries, reversors,
first integrals, or saddle connections,
then these have natural implications on the Melnikov displacement.
These claims are the subject of proposition~\ref{pro:geometric}.

We recall that a diffeomorphism $f:M \to M$ is \emph{symmetric} when there
exists a diffeomorphism $s:M \to M$ such that $f \circ s = s \circ f$,
and then $s$ is called a \emph{symmetry} of the map $f$.
Analogously, $f$ is \emph{reversible} when there exists a diffeomorphism
$r:M \to M$ such that $f \circ r = r \circ f^{-1}$,
and then $r$ is called a \emph{reversor} of the map $f$.
In many applications, $r$ is an involution, $r^2 = \Identity$, though
this need not be the case~\cite{GomezM04},
Finally, a function $I: M \to \Rset$ is a first integral of $f$ when $I \circ f=I$.

\begin{pro}\label{pro:geometric}
Let $f_\epsilon$ be a smooth family of diffeomorphisms verifying the
assumptions of definition~\ref{def:melnikov} and
let $\mathcal{D}$ be its Melnikov displacement.
\begin{enumerate}
\item\label{item:h0}
Given any smooth family of changes of variables $h_\epsilon:M \to M$,
the family
$\tilde{f}_\epsilon = h_\epsilon^{-1}\circ f_\epsilon \circ h_\epsilon$
also verifies the assumptions of definition~\ref{def:melnikov},
its Melnikov displacement $\tilde{\mathcal{D}}$ is defined on the saddle
connection $\tilde{\Sigma} \equiv h_0^{-1}(\Sigma)$, and
\begin{equation}\label{eq:h0}
\tilde{\mathcal{D}} = h_0^\ast \mathcal{D} \; .
\end{equation}
\item
The Melnikov displacement is invariant by the
pullback and the push-forward of the unperturbed map.
That is,
\begin{equation}\label{eq:f0}
f_0^\ast \mathcal{D} = (f_0)_\ast \mathcal{D} = \mathcal{D} \; .
\end{equation}
\item
If $f_\epsilon$ has a smooth family of:
\begin{enumerate}
\item
Symmetries $s_\epsilon:M \to M$ such that $s_0(\Sigma)=\Sigma$, then
\begin{equation}\label{eq:s0}
s_0^\ast \mathcal{D} = (s_0)_\ast \mathcal{D} = \mathcal{D} \;.
\end{equation}
\item
Reversors $r_\epsilon: M\to M$ such that $r_0(\Sigma)=\Sigma$, then
\begin{equation}\label{eq:r0}
r_0^\ast \mathcal{D} = (r_0)_\ast \mathcal{D} = -\mathcal{D} \;.
\end{equation}
If, in addition, $r_0$ is an involution and its fixed set
$R_0\equiv \{ \xi \in M : r_0(\xi)=\xi \}$ is a submanifold
that intersects $\Sigma$ transversely at some point $\xi_0$, then
$\mathcal{D}(\xi_0)=0$.
\item
First integrals $I_\epsilon: M \to \Rset$ such that
$A_\epsilon \cup B_\epsilon \subset I_\epsilon^{-1}(\{0\})$, then
\begin{equation}\label{eq:I0}
\Interior{\mathcal{D}} \rmd I_0 \equiv \rmd I_0(\mathcal{D}) = 0 \; .
\end{equation}
\item
Saddle connections $\Sigma_\epsilon \subset W^\un_\epsilon \cap W^\st_\epsilon$
such that $\Sigma_0=\Sigma$,
then $\mathcal{D} = 0$.
\end{enumerate}
\item
If there exists a vector field $\mathcal{X}: M \to TM$ such that its flow
commutes with $f_\epsilon$,
then the Lie derivative of the Melnikov displacement with respect to
$\mathcal{X}$ vanishes, that is,
\begin{equation}\label{eq:X}
L_\mathcal{X} \mathcal{D} = 0 \; .
\end{equation}
\end{enumerate}
\end{pro}

It is very important to stress that the above results make sense
as identities on the normal bundle of saddle connections.
The relation~\eref{eq:h0} makes sense because the pullback $h_0^\ast$
maps $T\Sigma$ onto $T\tilde{\Sigma}$,
whereas~\eref{eq:f0} makes sense because $T\Sigma$ is invariant by
the pullback $f_0^\ast$ and the push-forward $(f_0)_\ast$.
Similar arguments apply to~\eref{eq:s0} and \eref{eq:r0}.
The identities~\eref{eq:I0} and~\eref{eq:X} make sense because
$T\Sigma$ is contained in the kernel of the one-form $\rmd I_0$
and the vector field $\mathcal{X}$ is tangent to $\Sigma$, respectively.
The hypothesis
$A_\epsilon \cup B_\epsilon \subset I_\epsilon^{-1}(\{0\})$
means that $A_\epsilon$ and $B_\epsilon$ are contained in the same level set
of the first integrals, which can be assumed to be the zero level
without loss of generality.
This holds, for instance, in the transitive homoclinic case:
$B_\epsilon=A_\epsilon$ and $A_\epsilon$ is transitive.

\proof
We can write a geometric proof based on the definition of the
Melnikov displacement as a canonical section of the normal bundle,
or a computational proof using the formulae~\eref{eq:MelnikovSums}.
We follow the geometric approach.

(i)
The first claims are obvious.
For the last one, note that if
$\phi^{\st,\un}_\epsilon:W^{\st,\un}_0 \to W^{\st,\un}_\epsilon$
are any adapted deformations (for the family $f_\epsilon$),
then
$\tilde{\phi}^{\st,\un}_\epsilon \equiv
 h_\epsilon^{-1} \circ \phi^{\st,\un}_\epsilon \circ h_0$
are as well (for the family $\tilde{f}_\epsilon$).
Differentiating
$h_\epsilon \circ \tilde{\phi}^{\st,\un}_\epsilon =
 \phi^{\st,\un}_\epsilon \circ h_0$
with respect to $\epsilon$ yields
\[
\mathcal{H}_0 \circ h_0 +
{\rm D}h_0
\left. \frac{\partial}{\partial\epsilon}\right|_{\epsilon=0}
\tilde{\phi}^{\st,\un}_\epsilon =
\left. \frac{\partial}{\partial\epsilon}\right|_{\epsilon=0}
\phi^{\st,\un}_\epsilon \circ h_0
\]
where we used that
$\frac{\partial}{\partial \epsilon} h_\epsilon =
 \mathcal{H}_\epsilon \circ h_\epsilon$.
Hence,
$\mathcal{H}_0 \circ h_0 + {\rm D}h_0 \tilde{\mathcal{D}}^{\st,\un} =
 \mathcal{D}^{\st,\un} \circ h_0$,
so the difference $\mathcal{D} = \mathcal{D}^\un-\mathcal{D}^\st$ verifies the relation
$\mathcal{D} \circ h_0 = {\rm D} h_0 \tilde{\mathcal{D}}$,
which is equivalent to~\eref{eq:h0}.

(ii)
If we take $h_\epsilon=f_\epsilon$ in~\eref{item:h0}, then
$\tilde{f}_\epsilon =
 f_\epsilon^{-1} \circ f_\epsilon \circ f_\epsilon =
 f_\epsilon$ and
$\tilde{\Sigma}=f_0^{-1}(\Sigma)=\Sigma$.
Therefore, $\tilde{\mathcal{D}}=\mathcal{D}$
and~\eref{eq:f0} follows from~\eref{eq:h0}.

(iii.a)
If we take $h_\epsilon=s_\epsilon$ in~\eref{item:h0}, then
$\tilde{f}_\epsilon =
 s_\epsilon^{-1} \circ f_\epsilon \circ s_\epsilon = f_\epsilon$
and $\tilde{\Sigma}=s_0^{-1}(\Sigma)=\Sigma$.
Therefore, $\tilde{\mathcal{D}}=\mathcal{D}$
and~\eref{eq:s0} follows from~\eref{eq:h0}.

(iii.b)
If we take $h_\epsilon=r_\epsilon$ in~\eref{item:h0}, then
$\tilde{f}_\epsilon =
 r_\epsilon^{-1} \circ f_\epsilon \circ r_\epsilon = f_\epsilon^{-1}$
and $\tilde{\Sigma}=r_0^{-1}(\Sigma)=\Sigma$.
Moreover, a stable (respectively, unstable) manifold of a map
becomes an unstable (respectively, stable) for the inverse map.
Therefore, $\tilde{\mathcal{D}}=-\mathcal{D}$
and~\eref{eq:r0} follows from~\eref{eq:h0}.

Next,
we assume that $r_0$ is an involution whose fixed set $R_0$
intersects $\Sigma$ transversely at $\xi_0$.
That is, $r_0^2=\Identity$, $r_0(\xi_0)=\xi_0$,
and $T_{\xi_0} M = T_{\xi_0} \Sigma \oplus T_{\xi_0} R_0$.
We want to prove that $\mathcal{D}(\xi_0) = 0$ in the normal bundle $\nu(\Sigma)$,
or equivalently that $\mathcal{D}(\xi_0) \in T_{\xi_0} \Sigma$.

Since $r_0(\xi_0)=\xi_0$ and $r_0^2=\Identity$,
the square of the linear endomorphism
${\rm D}r_0(\xi_0):T_{\xi_0} M \to T_{\xi_0} M$ is the identity map.
This implies that ${\rm D}r_0(\xi_0)$ is diagonalizable and
its spectrum is contained in the set $\{-1,1\}$,
so $T_{\xi_0} M = E^+ \oplus E^-$,
where $E^\pm = \ker(Dr_0(\xi_0) \mp \Identity)$.
We claim that $E^+ = T_{\xi_0} R_0$ and $E^- = T_{\xi_0} \Sigma$.
The first claim follows from the fact that involutions are locally conjugate
to their linear parts at fixed points.
Since $T_{\xi_0} \Sigma$ is invariant under $Dr_0(\xi_0)$
and complementary to $T_{\xi_0} R_0 = E^+$ in $T_{\xi_0} M$,
we get the second claim.
Finally, the evaluation of~\eref{eq:r0} at the point $\xi=\xi_0$,
yields $Dr_0(\xi_0)\mathcal{D}(\xi_0) = -\mathcal{D}(r_0(\xi_0)) = -\mathcal{D}(\xi_0)$,
and so $\mathcal{D}(\xi_0) \in E^- = T_{\xi_0} \Sigma$.

(iii.c)
Let $I_\epsilon=I_0 + \epsilon I_1 + \Or(\epsilon^2)$.
Obviously, $W^{\un,\st}_\epsilon \subset I_\epsilon^{-1}(\{0\})$.
Differentiating $I_\epsilon \circ \phi^{\st,\un}_\epsilon = 0$
with respect to $\epsilon$ yields $\rmd I_0(\mathcal{D}^{\st,\un}) = -I_1$.
Thus,
$\mathcal{D} = \mathcal{D}^\un-\mathcal{D}^\st$ verifies~\eref{eq:I0}.

(iii.d)
In this case, we can take $\phi^\st_\epsilon=\phi^\un_\epsilon$, and so
$\mathcal{D} = \mathcal{D}^\un-\mathcal{D}^\st$ vanishes on
$\Sigma=\Sigma_0$.

(iv)
Let $\phi^t: M \to M$ be the flow of the vector field:
$\frac{\partial}{\partial t} \phi^t = \mathcal{X} \circ \phi^t$
and $\phi^0=\Identity$.
Thus, given any $t$,
the (constant) family $s_\epsilon \equiv \phi^t$ is a smooth family of
symmetries of $f_\epsilon$.
Next, using relation~\eref{eq:s0},
we get that $(\phi^t)^\ast \mathcal{D} = \mathcal{D}$ for any $t$,
and~\eref{eq:X} follows by definition of the Lie derivative.
\qed

These results have been extensively used in the literature. The
invariance of Melnikov objects under the unperturbed map gives
rise to periodicities when suitable coordinates are used; examples
can be found in~\cite{GlasserPB89,DelshamsR96,Levallois97}.
Item~(iii.d) implies a simple splitting criterion: if the Melnikov
displacement does not vanish identically, the separatrix
splits~\cite{DelshamsR96,RamirezRos06}. Upper bounds on the number
of uniform first integrals of the family $f_\epsilon$ can be
deduced from item~(iii.c), see~\cite{MeletlidouI94,WodnarIM99}.
This result has also been used in to establish necessary and
sufficient conditions for uniform integrability of analytic,
exact-symplectic maps~\cite{Llave96}. Symmetries have also been
extensively used, for example to improve the lower bound obtained
by Morse theory for the number of critical points of some Melnikov
potentials~\cite{DelshamsR97} (We will discuss Melnikov potentials
in \S\ref{sec:exactSymp}.). Relation~\eref{eq:X} is similar to
Noether's theorem, since the existence of a continuous
symmetry---the flow of the vector field $\mathcal{X}$---implies a
conservation law for the Melnikov displacement. Finally, fixed
sets of reversors can be used to guarantee the existence of
heteroclinic points and zeros of Melnikov functions; this is an
old trick, see~\cite{Devaney76}.

In the classical Melnikov method, one uses simple zeros of a Melnikov function
to predict the transverse intersection of the invariant manifolds.
We next show that this result also holds for the Melnikov displacement.

\begin{teo}
If $\xi_0$ is a simple zero of the Melnikov displacement~\eref{eq:melnikovDisp1},
then the perturbed invariant manifolds $W^\un_\epsilon$ and $W^\st_\epsilon$
intersect transversely at some point $\xi_\epsilon = \xi_0 + \Or(\epsilon)$
for small enough $\epsilon$.
\end{teo}

\begin{figure}
[ptb]
\begin{center}
\includegraphics[height=2in]{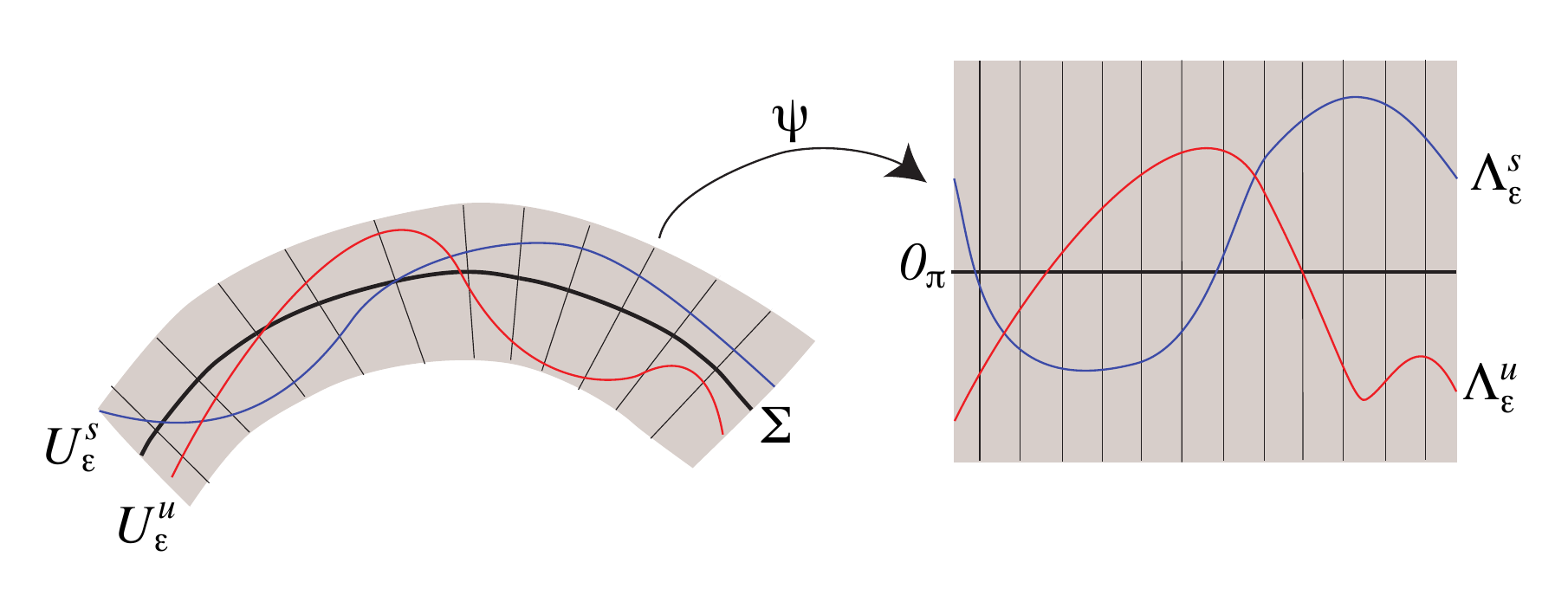}
\caption{A tubular neighbourhood that is diffeomorphic to the normal bundle
$\pi:\nu(\Sigma)\rightarrow\Sigma$.}
\label{fig:tub}
\end{center}
\end{figure}

\proof
By definition, the saddle connection $\Sigma$ is an $f_0$-invariant
submanifold of $M$.
Let $\pi:\nu(\Sigma)\rightarrow \Sigma$ be the projection of
the normal bundle onto $\Sigma$.
There is a tubular neighbourhood $N$ of $\Sigma$ that is diffeomorphic to
$\nu(\Sigma)$ via a diffeomorphism $\psi:N\rightarrow\nu(\Sigma)$,
as illustrated in~\fref{fig:tub}.
Since $N$ is an open neighbourhood of $\Sigma$ in $M$,
each deformation of $\Sigma$ occurs inside $N$, for $\epsilon$ small enough.

We note that under the diffeomorphism $\psi$, and for $\epsilon$ small
enough, deformations of $\Sigma$ can be thought as deformations
of the zero section, $0_{\pi}$, in $\nu(\Sigma)$.
Indeed, a deformation of $\Sigma$
can be parametrized by a section $\Sigma\rightarrow\nu(\Sigma)$.
Let $U_0\subset\Sigma$ be an open set that contains $\xi_0$.
Let $U_\epsilon^\st$ and $U_\epsilon^\un$ be
deformations of $U_0$ such that
\[
U_\epsilon^\st \subset W^\st_\epsilon \;,\qquad
U_\epsilon^\un \subset W^\un_\epsilon\;.
\]
Let $\Lambda_\epsilon^\st = \psi\left(U_\epsilon^\st\right)$ and
$\Lambda_\epsilon^\un = \psi\left(U_\epsilon^\un\right)$
be the images in $\nu(\Sigma)$ of two deformations of the saddle connection,
corresponding to the images of the stable and unstable manifolds. We want to
show that $\Lambda_\epsilon^\st$ and $\Lambda_\epsilon^\un$
intersect transversely for $\epsilon$ small enough when $\mathcal{D}$ has a
simple zero.

We parametrize each manifold with a section $U_0 \rightarrow \nu(\Sigma)$.
That is, if $\epsilon$ is small,
then $\tilde{\phi}_\epsilon^{\st,\un}:U_0\rightarrow\nu(\Sigma)$ given by
\[
\tilde{\phi}_\epsilon^{\st,\un}(\xi)=\pi^{-1}(\xi)\cap\Lambda_\epsilon^{\st,\un}
\]
are sections of the normal bundle restricted to $U_0$. Notice that
the functions $\phi_\epsilon^{\st,\un}:U_0 \to N\subset M$
given by
\[
\phi_\epsilon^{\st,\un}=\psi^{-1}\circ\tilde{\phi}_\epsilon^{\st,\un}
\]
are adapted deformations of $U_0$ with images $U_\epsilon^{\st,\un}$.
In fact, both are perturbations of the zero section.
From the definition of displacements we have that
\[
\tilde{\phi}_\epsilon^{\st,\un}(\xi) =
\epsilon \tilde{\mathcal{D}}^{\st,\un}(\xi) + \Or(\epsilon^{2})
\]
where
\[
\tilde{\mathcal{D}}^{\st,\un}(\xi) = {\rm D} \psi(\xi)\mathcal{D}^{\st,\un}(\xi)
\]
and $\mathcal{D}^{\st,\un}(\xi)$, are the displacements of
$\phi_\epsilon^{\st,\un}$. Therefore, each manifold
$\Lambda_\epsilon^\st$ and $\Lambda _\epsilon^\un$ is the image of
$\tilde{\phi}_\epsilon^\st$ and $\tilde{\phi}_\epsilon^\un$ and
they intersect transversely at
$\tilde{\phi}_\epsilon^\st(\xi_0)=\tilde{\phi
}_\epsilon^\un(\xi_0)$ if and only if $\xi_0$ is a simple zero of
the section $\tilde{\phi}_\epsilon^\un-\tilde{\phi}_\epsilon^\st$.

Now we use the standard ``blow up'' argument so that the implicit function
theorem can be applied.
Let $\bar{\phi}_\epsilon^{\st,\un}:\Sigma \rightarrow\nu(\Sigma)$
be the section given by
\[
\bar{\phi}_\epsilon^{\st,\un}=
\cases{
\case{1}{\epsilon} \tilde{\phi}_\epsilon^{\st,\un} & for $\epsilon \neq 0$ \\
\tilde{\mathcal{D}}^{\st,\un} & for $\epsilon=0$ \\
}
\]
Notice that $\bar{\phi}_\epsilon^{\st,\un}$ is $C^{r-1}$
if $\tilde{\phi}_\epsilon^{\st,\un}$ is $C^{r}$, and moreover
that, when $\epsilon \neq 0$, $\xi_\epsilon$ is a simple zero of
$\tilde{\phi}_\epsilon^\un - \tilde{\phi}_\epsilon^\st$ if and only if
it is for $\bar{\phi}_\epsilon^\un-\bar{\phi}_\epsilon^\st$.
Finally, since
${\rm D} \psi(\xi)\mathcal{D}(\xi) =
\left(  \bar{\phi}_0^\un - \bar{\phi}_0^\st\right)(\xi)$
and
$\Lambda_\epsilon^\st$ and $\Lambda_\epsilon^\un$ are the images of
$\tilde{\phi}_\epsilon^\st$ and $\tilde{\phi}_\epsilon^\un$,
the implicit function theorem implies
that if $\xi_0$ is a simple zero of $\mathcal{D}$,
then $\Lambda_\epsilon^\st$ and $\Lambda_\epsilon^\un$
intersect transversely near $\xi_0$, for $\epsilon$ small enough.
Thus $W^\st_\epsilon$ and $W^\un_\epsilon$
intersect transversely near $\xi_0$, for $\epsilon$ small enough.
\qed

\subsection{Example: perturbed Suris map}\label{ssec:Suris}

As a simple example, we consider the generalized standard map
\begin{eqnarray}\label{eq:genstd}
f:\Tset\times \Rset \to \Tset \times \Rset \;, \qquad
f(x,y) =  \big( x + y - V'(x), y- V'(x)\big)
\end{eqnarray}
where $V:\Tset \to \Tset$ is a periodic potential.
It is easy to see that $f$ preserves area and orientation and that
its fixed points have the form $(x^\ast,0)$ where $V'(x^\ast)=0$.
Such fixed points are saddles if and only if $V''(x^\ast)<0$,
because $\tr({\rm D} f(x^\ast,0)) = 2 - V''(x^\ast)$.

Following McMillan~\cite{McMillan71},
we can find a generalized standard map with a saddle connection between
two saddle points if we choose a diffeomorphism
$c:\Rset \to \Rset$ such that $c(x+2)=c^{-1}(x+1)+1=c(x)+2$,
and let
\begin{equation}\label{eq:secondDiff}
V'(x) = 2x - c(x) - c^{-1}(x)\;.
\end{equation}
To see this, first note that, with this choice,
the force $V'(x)$ is periodic with period one.
Moreover, if $x^\ast$ is a hyperbolic fixed point of $c$
(that is, $c(x^\ast)=x^\ast$  and $0<c'(x^\ast)\neq 1$), then
\[
V''(x^\ast) =
2 - c'(x^\ast) - (c^{-1})'(x^\ast) =
2 - c'(x^\ast) - 1/c'(x^\ast) < 0\;.
\]
Thus, $(x^\ast,0)$ is a saddle fixed point.

\begin{figure}[tbh]
\begin{center}
\includegraphics[width = 3.5in] {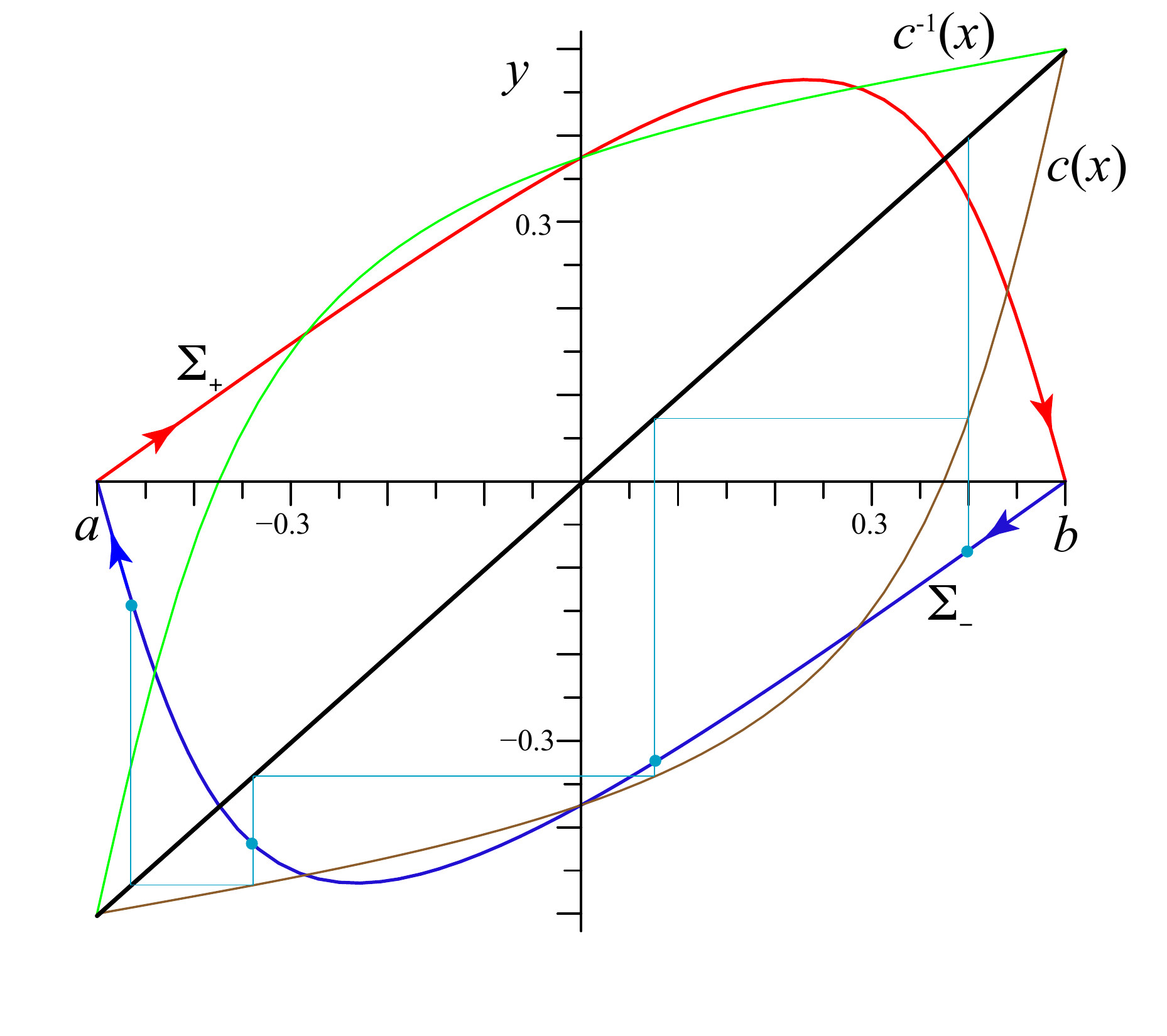}
\end{center}
\caption{Saddle connections $\Sigma_+$ and $\Sigma_-$ for the
Suris map with potential~\eref{eq:surispot} and the
diffeomorphisms $x \mapsto c^{\pm 1}(x)$ for $\mu=0.2$. Also shown
are images of a point on the saddle connection $\Sigma_{-}$,
obtained by iterating $x \mapsto c(x)$.}
\label{fig:surisManifolds}
\end{figure}

Moreover, the graphs of the functions $\chi_{\pm}(x) \equiv x-c^{\pm1}(x)$
are invariant and the dynamics on these sets is very simple:
\begin{equation}\label{eq:mcMIterates}
\eqalign{
f^k\big(x,\chi_-(x)\big) = \big(c^k(x),\chi_-(c^k(x))\big) \;, \cr
f^k\big(x,\chi_+(x)\big) = \big(c^{-k}(x),\chi_+(c^{-k}(x))\big)}
\end{equation}
for all $k\in\Zset$. These graphs contain saddle connections if we
choose a pair of neighbouring fixed points $a$ and $b$ of $c$, so
that $c$ has no fixed points in $(a,b)$. Suppose further that $a$
(respectively, $b$) is a stable (respectively, unstable) fixed
point of $c$, so that $\lim_{k\to +\infty} c^k(x)= a$ and
$\lim_{k\to +\infty} c^{-k}(x)=b$ for all $x\in(a,b)$. Then
$A=(a,0)$ and $B=(b,0)$ are saddle points of the map $f$, and
\[
\Sigma_{\pm} =
\left\{ (x,\chi_{\pm}(x))\in \Tset\times\Rset : x\in(a,b) \right\}
\]
are saddle connections between them:
$\Sigma_{-}\subset W^\st(A)\cap W^\un(B)$ and
$\Sigma_{+}\subset W^\un(A)\cap W^\st(B)$,
see~\fref{fig:surisManifolds}.

It is known that a generalized standard map with a potential of
the form~\eref{eq:secondDiff} is typically
nonintegrable~\cite{LomeliM03}. An integrable example, $f_0$, for
each $\mu\in(0,1)$ is obtained when the diffeomorphism $c$ is
given by
\begin{equation}\label{eq:cmu}
c(x) = c_\mu(x) =
\frac{2}{\pi}
\arctan\left(\frac{(\mu+1)\tan(\pi x/2)+(\mu-1)}
                  {(\mu-1)\tan(\pi x/2)+(\mu+1)}
\right)
\end{equation}
for $-1 \le x \le 1$. This function, when extended to $\Rset$ using
$c(x+2)=c(x)+2$, gives the period-one potential~\eref{eq:secondDiff}
\begin{equation}\label{eq:surispot}
V_0(x) =
\frac{2}{\pi} \int_0^{x}
\arctan\left(\frac{\delta \sin(2\pi t)}
                  {1+\delta\cos\ (2\pi t)}
       \right)
{\rm d} t \;, \qquad
\delta=\frac{(1-\mu)^2}{(1+\mu)^2} \;.
\end{equation}
A first integral of the map is
$I(x,y)=\cos\pi y+\delta\cos\pi(2x-y)$~\cite{Suris89,Meiss92,LomeliM00b}.

The diffeomorphism~\eref{eq:cmu} is conjugate to a M\"{o}bius
transformation, and it is easy to find an explicit formula for its iterations:
\begin{equation}\label{eq:cIterates}
{\left(c_\mu\right)}^k(x) = c_{\mu^k}(x) \;,\qquad \forall
k\in\Zset\;.
\end{equation}
The points $a=-1/2$ and $b=1/2$ are fixed points of $c$.
The point $a$ is stable and $b$ is unstable
because $c'(a) = \mu \in (0,1)$ and $c'(b) = 1/\mu > 1$.

\begin{figure}[tbh]
\begin{center}
\includegraphics[width = 3.5in]{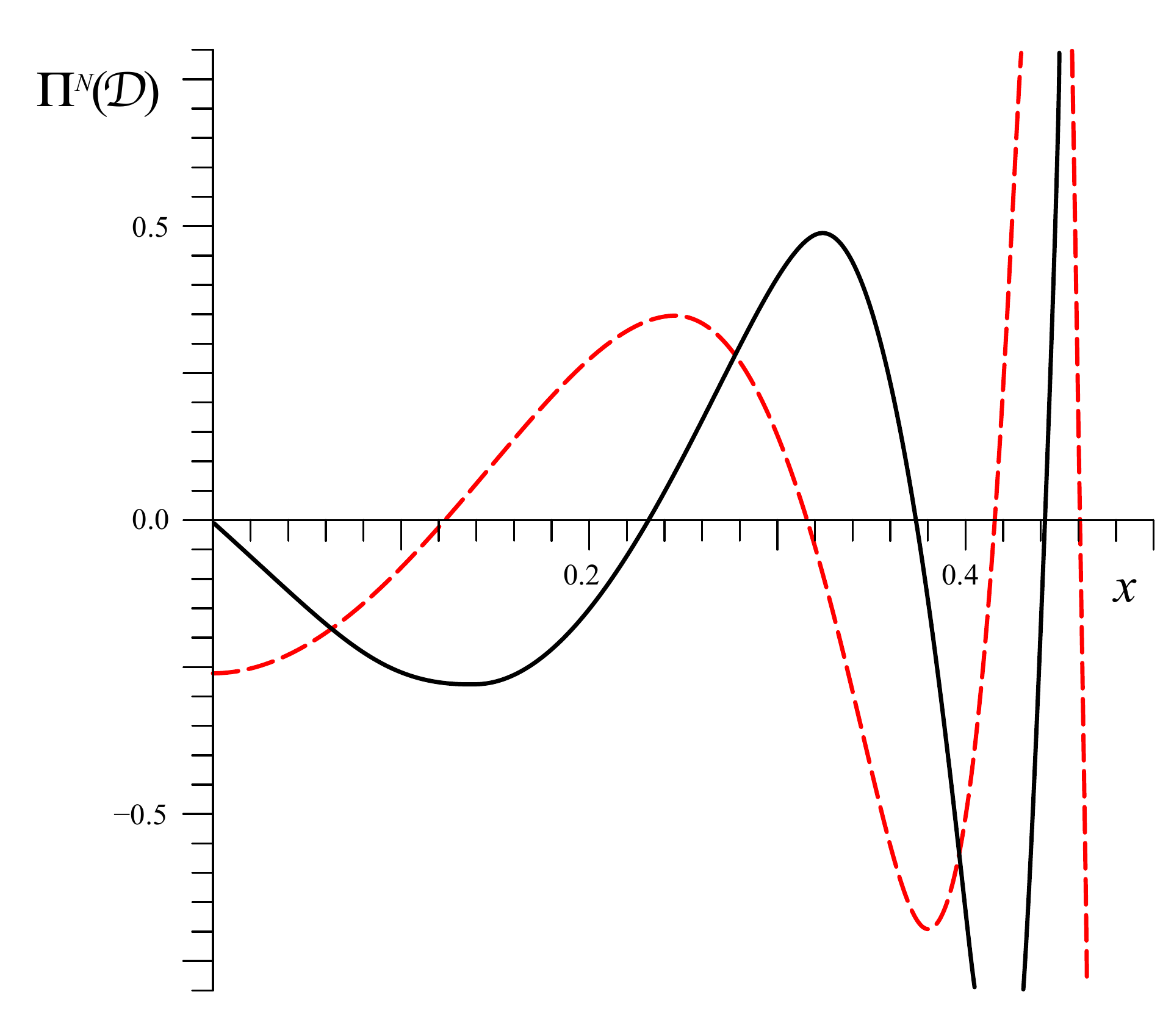}
\end{center}
\caption{Normal component of $\mathcal{D}$ for the Suris map on $\Sigma_{-}$
with $\mu= 0.2$. The solid curve is for $U(x) = -\cos(2\pi x)/2\pi$
and the dashed curve is for $U(x) = \sin(2\pi x)/2\pi$.}
\label{fig:surisDisp}
\end{figure}

We now perturb the integrable map by modifying
the potential~\eref{eq:surispot}:
\begin{equation}\label{eq:pertV}
V_\epsilon(x) = V_0(x) + \epsilon U(x) \;.
\end{equation}
We compute the Melnikov displacement using~\eref{eq:MelnikovSums}.
For this calculation we must first compute the vector field $\mathcal{F}_0$,
which for the generalized standard map~\eref{eq:genstd}
with potential~\eref{eq:pertV} is
\[
\mathcal{F}_0(x,y) = -U'(x-y)  (1,1)^T\;.
\]
Using the relations~\eref{eq:mcMIterates} and~\eref{eq:cIterates},
the series~\eref{eq:MelnikovSums} for the displacement $\mathcal{D}$
is easily computed. However, in order to ensure convergence of
the series, we must take into account the fact that it is only
the normal component of the displacement that is desired;
indeed, the iteration of the tangential component is not bounded.
Let $\Pi^N :T_\Sigma M \to \nu(\Sigma)$ be the canonical projection
onto the normal bundle~\eref{eq:normalbundle}.
Due to the fact that $\Sigma$ is invariant under $f_0$,
we have that $\Pi^N \circ f_0^\ast = f_0^\ast \circ \Pi^N$.
To avoid numerical errors,
we project each term in the sum~\eref{eq:MelnikovSums}.
We display in~\fref{fig:surisDisp} the projection $\Pi^N(\mathcal{D})$
as a function of $x$ along $\Sigma_{-}$,
for two perturbative potentials $U(x)$.

\section{Comparison with classical methods}\label{sec:classical}

The Melnikov displacement~\eref{eq:melnikovDisp1} generalizes the classical
methods used to detect the splitting of separatrices due to Poincar\'{e} and
Melnikov.

\subsection{Poincar\'{e} method}

Poincar\'{e}'s method~\cite{Poincare99} is based on the existence of first
integrals, such as an energy function, for the unperturbed system.
Assume that $f_0$ has a saddle connection $\Sigma$ between
a pair of hyperbolic fixed points $a$ and $b$.
For the simplest case, $\Sigma$ has codimension one, and
a single first integral suffices: $I\circ f_0=I$.
Saddle connections with a higher codimension can be treated in
a similar way if there are sufficiently many integrals. The splitting between
the stable and unstable manifolds is measured by the rate of change in the first
integral $I$ with $\epsilon$ on the saddle connection.
That is, we define the Poincar\'{e} function $M_{I}:\Sigma \rightarrow\Rset$ by
\begin{equation}\label{eq:classicP}
M_{I}=\left.  \frac{\partial}{\partial\epsilon}\right|  _{\epsilon=0}
\left(  I\circ\phi_\epsilon^\un-I\circ\phi_\epsilon^\st\right)
\end{equation}
where $\phi_\epsilon^\un$ and $\phi_\epsilon^\st$ are deformations adapted
to the perturbed invariant manifolds.

\begin{pro}
Under the above assumptions,
\[
M_{I}=\Interior{\mathcal{D}} {\rm d} I \equiv {\rm d} I(\mathcal{D})
\]
where $\mathcal{D}$ is the Melnikov displacement~\eref{eq:melnikovDisp1}.
\end{pro}

The proof is a simple computation.
The term ${\rm d}I(\mathcal{D})$ makes sense because
the first integral is constant on the saddle connection,
and so the tangent space $T\Sigma$ in contained in the kernel of ${\rm d}I$.
If ${\rm d} I|_{\Sigma}$ is nondegenerate, then a point $\xi_0\in\Sigma$ is a
simple zero of $M_{I}$ if and only if it is a simple zero of $\mathcal{D}$. In
this case, the perturbed invariant manifolds intersect transversely near
$\xi_0$ for $\epsilon$ small enough.
Therefore, the Melnikov displacement
$\mathcal{D}$ generalizes the estimates of splitting in the Poincar\'{e} style.

\subsection{Melnikov method}\label{ssec:Melnikovmethod}

The classical Melnikov method~\cite{Melnikov63} is based on estimating the
movement of a manifold in a direction {\emph normal} to the separatrix.
To define the normal, the phase space $M$ is assumed to have a Riemannian
inner product $\langle \cdot,\cdot \rangle$.
As before, assume for simplicity that $\Sigma$ is a saddle connection of
codimension one.
The appropriate normal to $\Sigma$ is called an adapted normal vector field.

\begin{defi}[Adapted Normal Vector Field~\cite{LomeliM00a}]\label{def:aniv}
Let $\Sigma$ be a $f_0$-invariant submanifold of codimension one.
A vector field $\eta:\Sigma \to T_\Sigma M$ is adapted normal
when
\begin{enumerate}
\item
$\eta$ is nondegenerate: $\eta(\xi) \neq0$, for all $\xi\in \Sigma$,
\item
$\eta$ is normal:
$\langle \eta(\xi) , v \rangle = 0$, for all $\xi\in\Sigma$ and $v \in T_\xi\Sigma$; and
\item
$\eta$ is invariant:
$f_0^\ast \langle \eta , \mathcal{Y} \rangle =
 \langle \eta , f_0^\ast \mathcal{Y} \rangle$,
for all vector field $\mathcal{Y}:\Sigma \rightarrow T_\Sigma M$.
\end{enumerate}
\end{defi}

Using this notation,
the classical Melnikov function $M_\eta:\Sigma \rightarrow\Rset$ is
\begin{equation}\label{eq:melnikovEta1}
M_\eta =
\left.  \frac{\partial}{\partial\epsilon}\right|  _{\epsilon=0}
\langle\eta,\phi_\epsilon^\un-\phi_\epsilon^\st \rangle
\end{equation}
where $\phi_\epsilon^\un$ and $\phi_\epsilon^\st$ are deformations
adapted to the perturbed invariant manifolds.
Consequently,
$M_\eta$ is related to the Melnikov displacement~\eref{eq:melnikovDisp1} by
\begin{equation}\label{eq:melnikovEta2}
M_\eta = \langle\eta,\mathcal{D} \rangle\;.
\end{equation}
Since $\eta$ is nondegenerate,
$M_\eta$ and $\mathcal{D}$ have the same simple zeros.
We also note that~\eref{eq:melnikovEta2} makes sense because $\eta$
is normal, and so the component of $\mathcal{D}$ in the tangent space $T\Sigma$
does not play any role.

\begin{remark}
Adapted normal vector fields can be obtained
as gradients of nondegenerate first integrals.
Let $I$ be a smooth first integral of $f_0$ such that
its gradient $\nabla I$ does not vanish on $\Sigma$.
Given a Riemannian structure, the gradient is the unique vector field such that
$\Interior{\mathcal{Y}} {\rm d} I = \langle \nabla I , \mathcal{Y} \rangle$
for any vector field $\mathcal{Y}$ on $M$.
Therefore,
$f_0^\ast \langle \nabla I , \mathcal{Y} \rangle =
 \langle \nabla(I\circ f_0) , f_0^\ast \mathcal{Y} \rangle$,
and $\eta=\nabla I$ is an adapted normal vector field.
Obviously,
the Poincar\'e function~\eref{eq:classicP} and
the Melnikov function~\eref{eq:melnikovEta1}
coincide when $\eta=\nabla I$.
\end{remark}

\section{Exact symplectic maps}\label{sec:symp}

\subsection{Basic results}

In this section,
we will see how the Melnikov displacement~\eref{eq:melnikovDisp1}
generalizes previous theories developed for hyperbolic fixed points of
exact symplectic maps.
We will also show why general normally hyperbolic invariant manifolds
can not be studied in the same way.

A $2n$-dimensional manifold $M$ is \emph{exact symplectic} when it
admits a nondegenerate two-form $\omega$ such that
$\omega=-{\rm d} \lambda$ for some Liouville one-form $\lambda$.
The typical example of an exact symplectic manifold is provided by
a cotangent bundle $M=T^\ast Q$, together with the canonical forms
$\omega_0 =  \rmd x \wedge \rmd y$ and $\lambda_0 = y \rmd x$, in
cotangent coordinates $(x,y)$.

A map $f:M \to M$ is \emph{exact symplectic} if
$\oint_\gamma \lambda = \oint_{f(\gamma)} \lambda$ for any closed
path $\gamma \subset M$ or, equivalently,
if there exists a \emph{generating} (or \emph{primitive}) \emph{function}
$S:M \to \Rset$ such that $f^\ast \lambda - \lambda = {\rm d} S$.
In particular, an exact symplectic map is \emph{symplectic}:
$f^\ast \omega = \omega$.

A submanifold $N$ of $M$ is \emph{exact isotropic} if
$\oint_\gamma \lambda = \oint_{f(\gamma)} \lambda$ for any closed
path $\gamma \subset N$ or, equivalently,
if there exists a generating function $L:N \to \Rset$ such that
$j^\ast_N \lambda = {\rm d} L$.
Here,
$j_N:N \hookrightarrow  M$ denotes the natural inclusion map.
In particular, an exact isotropic submanifold is \emph{isotropic}:
$j^\ast_N \omega = 0$.
The maximal dimension of an isotropic submanifold is $n$, and
when the dimension is $n$, the submanifolds is called \emph{Lagrangian}.

A vector field $\mathcal{F}: M \to TM$ is \emph{globally Hamiltonian} if
there exists a \emph{Hamiltonian} function  $H: M \to \Rset$ such that
\[
\Interior{\mathcal{F}} \omega = {\rm d} H \;.
\]

We stress two key properties of the symplectic case.
First,
the generator of a family of exact symplectic maps is globally Hamiltonian
and there exists a simple relation between this Hamiltonian and
the generating function of the maps.
Second,
the stable and unstable invariant manifolds of
a connected normally hyperbolic invariant submanifold $A$ of
an (exact) symplectic map are (exact) isotropic if and only if
$A$ is a fixed point,
in which case they are Lagrangian (that is, $n$-dimensional).
The second property is an obstruction to develop a symplectic version
of our canonical Melnikov theory for general normally hyperbolic
invariant submanifolds, we shall do it just for fixed points.

These properties are well-known, but we prove both for
completeness.

\begin{pro}\label{pro:Hamiltonian}
Let $f_\epsilon$ be a family of exact symplectic maps with generating
function $S_\epsilon$ and generating vector field $\mathcal{F}_\epsilon$.
Then, $\mathcal{F}_\epsilon$ is globally Hamiltonian with Hamiltonian
\begin{equation}\label{eq:Hamiltonian}
H_\epsilon =
\lambda(\mathcal{F}_\epsilon)-
\frac{\partial S_\epsilon}{\partial\epsilon}\circ f_\epsilon^{-1}\;.
\end{equation}
\end{pro}

\proof
By definition,
$\case{\partial}{\partial\epsilon}f_\epsilon = \mathcal{F}_\epsilon \circ f_\epsilon$,
so the Lie derivative with respect to $\mathcal{F}_\epsilon$ is
$L_{\mathcal{F}_\epsilon} \lambda=
 (f_\epsilon^\ast)^{-1} \case{\partial}{\partial \epsilon} (f_{\epsilon}^\ast\lambda)$.
Using Cartan's formula:
$L_{\mathcal{X}} \lambda =
 \Interior{\mathcal{X}} {\rm d} \lambda + {\rm d} \Interior{\mathcal{X}} \lambda$,
and taking the derivative with respect to $\epsilon$ of the relation
${\rm d} S_\epsilon = f_\epsilon^\ast \lambda - \lambda$, we get
\[
{\rm d} \frac{\partial S_\epsilon}{\partial\epsilon} =
\frac{\partial}{\partial\epsilon}f_\epsilon^\ast\lambda=
f_\epsilon^\ast L_{\mathcal{F}_\epsilon}\lambda =
f_\epsilon^\ast
\left( \Interior{\mathcal{F}_\epsilon} {\rm d} \lambda +
       {\rm d} \Interior{\mathcal{F}_\epsilon}\lambda\right)\;.
\]
Rearranging this yields
\[
\Interior{\mathcal{F}_\epsilon} {\rm d} \lambda =
\left(  f_\epsilon^{-1}\right)^\ast
{\rm d} \frac{\partial S_\epsilon}{\partial\epsilon} - {\rm d}
\Interior{\mathcal{F}_\epsilon}\lambda =
{\rm d} \left(
\frac{\partial S_\epsilon}{\partial\epsilon}\circ f_\epsilon^{-1}-
\Interior{\mathcal{F}_\epsilon}\lambda
\right)  \;.
\]
Finally, $\Interior{\mathcal{F}_\epsilon}\lambda=\lambda(\mathcal{F}_\epsilon)$,
since $\lambda$ is a one-form.
\qed

\begin{pro}\label{pro:isotropic}
Let $f:M \to M$ be a diffeomorphism with a connected, normally hyperbolic,
invariant manifold $A$. If $f$ is (exact) symplectic,
the stable and unstable invariant manifolds $W^{\st,\un}=W^{\st,\un}(A,f)$
are (exact) isotropic if and only if
%$\dim A=0$; that is, if and only if
$A$ is a hyperbolic fixed point.
Moreover, in this case the submanifolds $W^{\st,\un}$ are (exact) Lagrangian.
\end{pro}
\proof

Since $A$ is normally hyperbolic, \eref{eq:splitting} implies
that for each $a \in A$, $s = \dim E^\st_a$, $u =\dim E^\un_a$
and $c = \dim T_a A$ sum to $\dim T_a M = \dim M = 2n$. Since
$f$ is symplectic $s = u = n-c/2$. Therefore
$\dim W^\st = \dim W^\un = u + c = n+c/2$.
Suppose that $W^{\st,\un}$ are isotopic; then since isotopic manifolds
have maximum dimension $n$, we must have $c = 0$.
Consequently, $\dim A = 0$ and since $A$ is connected it is a hyperbolic fixed point.
%
%Let $s=\dim W^\st-\dim A$ and $u=\dim W^\un - \dim A$.
%Using the splitting~\eref{eq:splitting} in any point $a\in A$,
%we get that
%\begin{eqnarray*}
%2 n = \dim M & = & \dim T_a M \\
%& = &
%\dim T_a A + \dim E^\st_a + \dim E^\un_a \\
%& = &
%\dim A + (\dim W^\st - \dim A) + (\dim W^\un - \dim A) \\
%& = &
%\dim W^\st + \dim W^\un - \dim A \;.
%\end{eqnarray*}
%Moreover,
%from the symplectic character of the map $f$, we know that the number
%of stable and unstable directions coincide:
%$s=u$ and $\dim W^\st = \dim W^\un$.
%Therefore, $\dim W^{\st,\un}= n + (\dim A)/2$,
%so that $W^{\st,\un}$ can not be isotropic if $\dim A > 0$.

Conversely, assume that $\dim A=0$, so that $A=\{a\}$ for some hyperbolic
fixed point $a$ and $\dim W^{\st,\un} = n$. To prove that $W^\st$ is
Lagrangian, take any two vectors $u,v\in T_\xi W^\st$.
We know that ${\rm D}f^k u$ and ${\rm D}f^k v$ tend to zero as $k\to +\infty$,
since the stable directions are uniformly contracted.
Since $f$ preserves $\omega$, we have
\[
\omega(u,v) = (f^k)^\ast \omega(u,v) =
\omega({\rm D} f^k u , {\rm D} f^k v) \longrightarrow \omega(0,0) = 0
\]
as $k\to +\infty$.

If, in addition, $f$ is exact symplectic, then
for every closed loop $\gamma$, $\oint_\gamma \lambda = \oint_{f(\gamma)} \lambda$.
Now suppose that $\gamma \subset W^\st$, so that $f^k(\gamma) \to \{a\}$ as
$k \to +\infty$. Then
\[
\oint_\gamma \lambda = \oint_{f^k(\gamma)} \lambda \longrightarrow
\oint_a \lambda = 0.
\]
Finally, the (exact) Lagrangian character of $W^\un$ follows from the
fact that it is the stable manifold for $f^{-1}$.
\qed

\subsection{Melnikov method for hyperbolic fixed points of exact
symplectic maps}\label{sec:exactSymp}

Let $f_\epsilon:M \to M$ be a family of exact symplectic maps
such that $f_0$ has an exact Lagrangian saddle connection $\Lambda$
between two hyperbolic fixed points $a$ and $b$. Note
that in this case that the stable and unstable manifolds
are as smooth as the map $f_\epsilon$. We assume that $H_0(a)=H_0(b)$,
where $H_\epsilon$ is the Hamiltonian~\eref{eq:Hamiltonian}.
Without loss of generality, we can assume that $H_0(a)=H_0(b)=0$.

The natural measure of splitting for this case is a real valued function
$L$, the \emph{Melnikov potential},
whose derivative measures the splitting~\cite{DelshamsR97}.
In other words, nondegenerate critical points of the Melnikov potential
predict transverse splitting.
In this subsection,
we shall find the relation between the one form $dL$ introduced
by~\cite{Roy06} and the Melnikov displacement.

\begin{defi}[Melnikov Potential]
For a saddle connection $\Lambda$ and Melnikov displacement
$\mathcal{D}:\Lambda \to \nu(\Lambda)$,
the function $L:\Lambda\rightarrow\Rset$
implicitly defined by
\begin{equation}\label{eq:MelnikovPotential1}
{\rm d} L = j^\ast_\Lambda (\Interior{\mathcal{D}} \omega)\; .
\end{equation}
is the Melnikov potential.\footnote{
%%%%%
For simplicity, henceforth we will write this relation as
$\Interior{\mathcal{D}} \omega = {\rm d} L$.
%%%%%
}
\end{defi}

We will show next that $L$ is indeed defined by~\eref{eq:MelnikovPotential1}
and that its critical points correspond to zeros of the displacement.

\begin{pro}
The pullback of the one-form $\Interior{\mathcal{D}}\omega$
to the saddle connection $\Lambda$ is well-defined and exact.
In particular, there exists a function $L:\Lambda \to \Rset$,
determined uniquely up to additive constants that
obeys~\eref{eq:MelnikovPotential1}.

Moreover, the set of simple zeros of the Melnikov displacement $\mathcal{D}$
coincides with the set of nondegenerate critical points of the function $L$.
\end{pro}

\begin{remark}\label{rem:isotropic}
Let $N$ be a submanifold of $M$ and
$\mathcal{X}:N \to \nu(N)$ a section of its normal bundle.
Then the pullback of the one-form $\Interior{\mathcal{X}}\omega$
to the submanifold $N$ is \emph{well-defined} if and only if $N$ is isotropic.
This has to do with the fact that
$j^\ast_N (\Interior{\mathcal{X}} \omega)$ is well-defined if and only if
\[
\omega(\mathcal{X} + \mathcal{Y}, \mathcal{Z} ) =
\omega(\mathcal{X},\mathcal{Z})
\]
for any vector fields $\mathcal{X}: N \to T_N M$ and
$\mathcal{Y},\mathcal{Z} : N \to T N$.
\end{remark}

\proof
The pullback is well-defined because $\Lambda$ is Lagrangian,
see remark~\ref{rem:isotropic}.
With regard to the exactness, it suffices to prove that
$\oint_\gamma j^\ast_\Lambda(\Interior{\mathcal{D}} \omega) = 0$ for any
closed path $\gamma \subset \Lambda$.
Let $\gamma$ a closed path contained in the saddle connection, then
\[
\oint_\gamma j^\ast_\Lambda(\Interior{\mathcal{D}} \omega) =
\oint_\gamma j^\ast_\Lambda ( \Interior{f_0^\ast \mathcal{D}} (f_0^\ast \omega) ) =
\oint_{f_0(\gamma)} j^\ast_\Lambda(\Interior{\mathcal{D}} \omega)
\]
where, since $f_0$ is symplectic, the
Melnikov displacement is invariant under the pullback of $f_0$,
and the saddle connection is invariant under $f_0$.
Finally, since $\gamma \in W^\st_0(b,f_0)$, we obtain that
\[
\oint_\gamma j^\ast_\Lambda (\Interior{\mathcal{D}} \omega) =
\oint_{f_0^k(\gamma)} j^\ast_\Lambda (\Interior{\mathcal{D}} \omega)
\longrightarrow
\oint_{b} j^\ast_\Lambda (\Interior{\mathcal{D}} \omega) = 0
\]
as $k \to +\infty$.

The equivalence between simple zeros of $\mathcal{D}$ and nondegenerate
critical points of $L$ follows from the Lagrangian character of the saddle
connection $\Lambda$.
\qed

An explicit series for $L$ can be obtained using the
Hamiltonian~\eref{eq:Hamiltonian}.
\begin{cor}
The Melnikov potential~\eref{eq:MelnikovPotential1}
is given by the absolutely convergent series
\begin{equation}\label{eq:MelnikovPotential2}
L = \sum_{k\in\Zset} H_0 \circ f_0^k\;.
\end{equation}
\end{cor}

\proof
Since $f_0$ is symplectic, and the generator $\mathcal{F}_0$ is globally
Hamiltonian with Hamiltonian $H_0$:
$\Interior{\mathcal{F}_0}\omega = {\rm d} H_0$,
we deduce that
\[
\Interior{\mathcal{D}} \omega =
\sum_{k\in\Zset} \Interior{(f_0^\ast)^{k}\mathcal{F}_0} \omega =
\sum_{k\in\Zset} (f_0^\ast)^{k} \Interior{\mathcal{F}_0} \omega =
\sum_{k\in\Zset} {\rm d} (H_0 \circ f_0^k) =
{\rm d} L\;.
\]
The series converges absolutely because $H_0(a)=H_0(b)=0$,
$\lim_{k\to -\infty} f_0^k(\xi) = a$, and
$\lim_{k\to +\infty} f_0^k(\xi) = b$.
In fact, it converges at a geometric rate.
\qed

While the Melnikov potential has been used many times for
exact symplectic, twist, and Hamiltonian maps,
the formulation give here, using the power of deformation theory,
is more elegant.

The Melnikov potential introduced here is identical to the one defined
in~\cite{DelshamsR97}.
This can be checked by direct comparison of the
formula~\eref{eq:MelnikovPotential2} with the formula~(2.7) of the cited paper,
using~\eref{eq:Hamiltonian} to express $H_0$ in terms of
the derivative of the generating function $S_\epsilon$ at $\epsilon=0$.

We also note that it is impossible to define a ``Melnikov potential"
on the saddle connection of normally hyperbolic invariant manifolds
with nonzero dimension,
because then the saddle connection is not isotropic
(proposition~\ref{pro:isotropic})
and so the identity~\eref{eq:MelnikovPotential1} makes no sense
(remark~\ref{rem:isotropic}). Nevertheless, even in this case
the Melnikov displacement is defined.

\subsection{Area-preserving maps}\label{sec:APMaps}

In this subsection, we restrict to the two-dimensional case in order
to show in a simple way that the Melnikov potential and the classical Melnikov
function are transparently related. Moreover, we will see that in some cases
there exist geometric obstructions to the non-vanishing of Melnikov functions.
These points are most easily seen by choosing a special time-like parametrization
of the saddle connection.

We consider the standard symplectic structure on the plane,
$(M,\omega)=(\Rset^2,{\rm d} x \wedge {\rm d} y)$,
and let  $J$ be the standard $2\times 2$ symplectic matrix:
$\omega(u,v)=u^T J v=\langle u,Jv \rangle$.

Suppose that $f_\epsilon:\Rset^2 \to \Rset^2$ is a family of diffeomorphisms
preserving area and orientation, and $H_\epsilon$ is the generating Hamiltonian
for $\mathcal{F}_\epsilon$, the generator for $f_\epsilon$.
We assume that the unperturbed map has a
saddle connection $\Sigma\subset W^\un(a)\cap W^\st(b)$ between two hyperbolic
fixed points $a$ and $b$ such that $H_0(a) = H_0(b) = 0$.
Note that the unperturbed map not need be integrable.

The key point is that, on a one-dimensional saddle connection,
there is  a parametrization $\alpha:\Rset \to \Sigma$ such that
\[
f_0(\alpha(t))=\alpha(t+1) \;,\qquad
\lim_{t\to -\infty}\alpha(t)=a \;,\qquad
\lim_{t\to +\infty}\alpha(t)=b\;.
\]

\begin{remark}
In many cases, such parametrizations can be expressed in terms of
elementary functions, and the Melnikov function can be explicitly
computed~\cite{GlasserPB89,DelshamsR96,Lomeli96}.
\end{remark}

Consequently,
$\alpha$ provides a diffeomorphism between the saddle connection $\Sigma$
and the real line, so that objects defined over $\Sigma$ can be considered
as depending on the real variable $t$.
Thus, for example, the Melnikov potential~\eref{eq:MelnikovPotential2}
can be replaced by $L \circ \alpha$ to become a function $L:\Rset \to \Rset$
given by
\begin{equation}\label{eq:LofT}
L(t) = \sum_{k\in\Zset} H_0(\alpha(t+k)) \;.
\end{equation}
Here we abuse the notation by not giving the function a new name.

Our goal is to show that the classical Melnikov function~\eref{eq:melnikovEta2},
or rather the composition $M \circ \alpha$, can be computed by differentiating
$L$:
\begin{equation}\label{eq:MofT}
M(t) = L'(t) \;.
\end{equation}
In addition we will show that these functions have the properties
\begin{itemize}
\item
Periodicity: $L(t+1)=L(t)$ and $M(t+1)=M(t)$.
\item In each fundamental domain
$[t,t+1)$, $M$ must vanish.
Indeed, $\int_{t}^{t+1} M(s) {\rm d}s= 0$ for any $t\in\Rset$.
(This property will be generalized to volume-preserving maps
 in the next section.)
\item Near each simple zero of $M(t)$ or nondegenerate critical point
of $L(t)$ there is transverse intersection of the stable and unstable manifolds.
\end{itemize}

Given~\eref{eq:LofT} and~\eref{eq:MofT},
the first two properties are obvious. Periodicity is simply a consequence of
the invariance of~\eref{eq:LofT} under $t \to t+1$; this invariance under the
unperturbed map is a property of all of the Melnikov objects introduced so far.
The second property is a simple consequence of periodicity and integration
of~\eref{eq:MofT}.

To show that $M(t)$ is actually the classical Melnikov
function~\eref{eq:melnikovEta2} we must construct an adapted
normal vector field $\eta$ on $\Lambda$, recall definition~\ref{def:aniv}.
We claim that, when thought of as a function of $t$,
such a vector field $\eta:\Rset \to \Rset^2$ is given by
\[
\eta(t) = J \alpha'(t) \;.
\]
This claim is proved in lemma~\ref{lem:alpha} at the end of this subsection.

The relation between $M$ and the classical Melnikov function~\eref{eq:melnikovEta2}
follows from a straightforward computation of the derivative using $J^2 = -I$:
\begin{eqnarray*}
L'(t)   &  = &
\sum_{k\in\Zset} \langle \nabla H_0(\alpha(t+k)),\alpha'(t+k) \rangle \\
& = &
\sum_{k\in\Zset} \langle J\alpha'(t+k),-J\nabla H_0(\alpha(t+k)) \rangle \\
& = &
\sum_{k\in\Zset} \langle \eta(t+k),\mathcal{F}_0(\alpha(t+k)) \rangle \; .
\end{eqnarray*}
Thus
\[
M(t) =
\sum_{k\in\Zset} \langle \eta(t+k),\mathcal{F}_0(\alpha(t+k)) \rangle
\]
which is the obvious form of~\eref{eq:melnikovEta2} under
the parametrization $\alpha$. This verifies~\eref{eq:MofT} and the final property.

To end this subsection,
it remains to prove the claim about the vector field $\eta$.

\begin{lem}\label{lem:alpha}
The vector field $\eta:\Sigma \to \Rset^2$ defined by
$\eta(\alpha(t))=J\alpha'(t)$ is an adapted normal vector field
on the saddle connection $\Sigma$.
\end{lem}

\proof
Since $f_0$ is symplectic, ${\rm D} f_0^T   J   {\rm D} f_0=J$.
Consequently,
\begin{eqnarray*}
J\alpha'(t+1)
& = &
{\rm D} f_0^{-1}(\alpha(t+1))^T   J
{\rm D} f_0^{-1}(\alpha(t+1))   \alpha'(t+1) \\
& = &
{\rm D} f_0^{-1}(\alpha(t+1))^T   J   \alpha'(t)\;.
\end{eqnarray*}
Thus, given any vector field $\mathcal{Y}:\Sigma\rightarrow\Rset^2$,
we have
\begin{eqnarray*}
f_0^\ast \langle \eta , \mathcal{Y} \rangle (\alpha(t))
& = &
\langle \eta(\alpha (t+1)) , \mathcal{Y}(\alpha(t+1)) \rangle \\
& = &
\langle J  \alpha'(t+1), \mathcal{Y}(\alpha(t+1))\rangle \\
& = &
\langle J  \alpha'(t), {\rm D} f_0^{-1}(\alpha(t+1))   \mathcal{Y}(\alpha(t+1))\rangle \\
& = &
\langle J  \alpha'(t), {\rm D} f_0^{-1}(f_0(\alpha(t))   \mathcal{Y}(f_0(\alpha(t))\rangle \\
& = &
\langle \eta , f_0^\ast \mathcal{Y} \rangle (\alpha(t))\;.
\end{eqnarray*}
Moreover, since $J$ is antisymmetric, $\langle\eta,\alpha'\rangle=0$.
Therefore, according to definition~\ref{def:aniv},
$\eta:\Sigma \to \Rset^2$ is an adapted normal vector field.
\qed

\section{Volume-preserving maps}\label{sec:volume}

For the case of a volume-preserving mapping with a codimension-one saddle
connection, an adapted normal field formulation of the Melnikov function also
applies~\cite{LomeliM00a}. Here we show how to relate this to the Melnikov
displacement~\eref{eq:melnikovDisp1}.

Let $f_\epsilon:M\rightarrow M$ be a family of volume-preserving
diffeomorphisms on an oriented $n$-dimensional manifold $M$ with volume form
$\Omega$ such that $f_0$ has a codimension-one saddle connection $\Sigma$
between two normally hyperbolic invariant sets $A$ and $B$.

We start with a simple lemma about the generator for $f_\epsilon$.

\begin{pro}\label{pro:exact}
Let $\mathcal{F}_\epsilon$ be the generator of a volume-preserving
smooth family $f_\epsilon$.
Then
\begin{enumerate}
\item
The divergence of $\mathcal{F}_\epsilon$ with respect to $\Omega$ is zero.
\item
The one-form $\Interior{\mathcal{F}_\epsilon}\Omega$ is closed.
\item
If $M$ is simply connected, then $\Interior{\mathcal{F}_\epsilon}\Omega$ is exact.
\end{enumerate}
\end{pro}

\proof
 From~\cite[Thm.~2.2.21]{AbrahamM78}, we have that $L_{\mathcal{F}_\epsilon} \Omega=0$.
Since the divergence is defined by
$\Divergence (\mathcal{F}_\epsilon) \Omega = L_{\mathcal{F}_\epsilon} \Omega$,
this implies that the divergence vanishes.
Moreover, since ${\rm d} \Omega= 0$ and
$L_{\mathcal{F}_\epsilon} \Omega =
 {\rm d} \Interior{\mathcal{F}_\epsilon} \Omega
 + \Interior{\mathcal{F}_\epsilon} {\rm d} \Omega$,
then ${\rm d} \Interior{\mathcal{F}_\epsilon} \Omega = 0$, implying (ii) and (iii).
\qed

%\begin{remark}
%The following method to generate families of volume-preserving is useful.
%Suppose $f_0:\Rset^n \to \Rset^n$ is a volume-preserving map
%with respect to the standard volume form. Then if
%$P:\Rset^n \to \Rset^n$ is a map whose differential is nilpotent
%then $f_\epsilon = (\Identity + \epsilon P) \circ f_0$ preserves volume,
%see~\cite{LomeliM00a}.
%\end{remark}

In order to find an invariant nondegenerate $(n-1)$-form from
any adapted normal field,
we assume that $M$ has a Riemannian metric $\langle\cdot,\cdot\rangle$.

\begin{pro}
If $\eta$ is an adapted vector field (cf. definition~\ref{def:aniv}),
then
\begin{equation}\label{eq:omegaEta}
\omega_\eta =
\frac{\Interior{\eta} \Omega}{\langle\eta,\eta\rangle}
\end{equation}
is a nondegenerate $(n-1)$-form on $\Sigma$ that is invariant under the
restriction $f_0|_\Sigma$.
\end{pro}

\proof
By definition $\eta$ is nonzero, so that $\omega_\eta$ is nondegenerate.
To prove that $f_0^\ast\omega_\eta=\omega_\eta$ on the saddle
connection $\Sigma$, we introduce the vector field
$\mathcal{Z}_\eta:\Sigma \to T \Sigma$ defined by
\[
\mathcal{Z}_\eta \equiv
\frac{f_0^\ast\eta}{\langle\eta,f_0^\ast\eta\rangle} -
\frac{\eta}{\langle\eta,\eta\rangle}\;.
\]
This vector field is tangent to $\Sigma$,
because $\langle \eta , \mathcal{Z}_\eta \rangle \equiv 0$.
Now, we compute the difference
\[
f_0^\ast\omega_\eta-\omega_\eta =
\frac{\Interior{f_0^\ast \eta} f_0^\ast\Omega}
     {f_0^\ast \langle \eta , \eta \rangle} -
\frac{\Interior{\eta} \Omega}{\langle \eta , \eta \rangle} =
\frac{\Interior{f_0^\ast\eta} \Omega}
     {\langle \eta , f_0^\ast \eta \rangle} -
\frac{\Interior{\eta} \Omega}{\langle \eta , \eta \rangle} =
\Interior{\mathcal{Z}_\eta} \Omega\;.
\]
Hence, it suffices to see that the $(n-1)$-form $\Interior{\mathcal{Z}_\eta}\Omega$
vanishes identically on the tangent space $T \Sigma$.
This follows from the fact that $\mathcal{Z}_\eta$ is tangent to $\Sigma$ and
$\dim \Sigma = n-1$.
\qed

The Melnikov function associated with $\eta$ is defined using $\omega_\eta$.

\begin{defi}[Volume-Preserving Melnikov Function]\label{def:java}
Let $\mathcal{D}$ be the Melnikov displacement~\eref{eq:melnikovDisp1}.
Given an adapted vector field $\eta$ on $\Sigma$,
we define the Melnikov function $M_\eta: \Sigma \to \Rset$ as the unique
$C^{r-1}$ function such that
\begin{equation}\label{eq:melnikovVolume}
M_\eta \omega_\eta =\Interior{\mathcal{D}} \Omega
\end{equation}
as $(n-1)$-forms on $\Sigma$.
\end{defi}

The previous definition first appeared in \cite{LomeliM00a}. Note
that $\Interior{\mathcal{D}} \Omega$ is an $(n-1)$-form, but it is
possible that it might be degenerate; if this were the case then
zeros of $M_\eta$ need not correspond to those of $\mathcal{D}$.
We will show next that this is not the case.

\begin{pro}\label{pro:java}
The Melnikov function $M_\eta$ is invariant under the map $f_0$
and $M_\eta = \langle \eta , \mathcal{D} \rangle$.
Moreover,
a point $\xi_0$ is a zero of $M_\eta$ if and only if it is a zero of $\mathcal{D}$.
\end{pro}

\proof
Using~\eref{eq:omegaEta}, we obtain
\[
\Interior{\mathcal{D}} \Omega - \langle \eta, \mathcal{D} \rangle \omega_\eta =
\Interior{v} \Omega
\]
where
$\displaystyle v\equiv\mathcal{D} -
 \frac{\langle\eta,\mathcal{D}\rangle}
      {\langle\eta,\eta\rangle}\eta$.
Since $v\in T\Sigma$,
we conclude that $\Interior{v} \Omega \equiv 0$,
as an $(n-1)$-form on $\Sigma$,
and thus $M_\eta = \langle \eta , \mathcal{D} \rangle$.
Since $\eta$ is an adapted field, and $\mathcal{D}$ is invariant,
$f_0^\ast \langle \eta , \mathcal{D} \rangle =
 \langle \eta , f_0^\ast \mathcal{D} \rangle =
 \langle \eta , \mathcal{D} \rangle$.
Therefore, $f_0^\ast M_\eta=M_\eta$.
Finally, $M_\eta(\xi_0)=0$ only when $\mathcal{D}(\xi_0)=0$,
since $\eta$ is nonzero and normal to $T\Sigma$,
and $\mathcal{D}$ is not in the tangent space.
\qed

We would like to show,
as we did for the area-preserving case in~\sref{sec:APMaps},
that the volume-preserving Melnikov function necessarily has zeros
on $\Sigma$.
To do this, we will show that the integral of $M_\eta$ with respect
to the measure $\omega_\eta$ is zero.
This can be accomplished by dividing the saddle connection into
pieces---fundamental domains---that are mapped into each other by $f_0$.

\begin{defi}[Proper Boundary]\label{def:guasa}
Let $A$ be a compact normally hyperbolic invariant manifold of a
diffeomorphism $f_0$ with stable manifold $W^\st(A)$.
A proper boundary,
$\gamma$, is a submanifold of $W^\st(A)$ that bounds an isolating neighbourhood
of $A$ in $W^\st(A)$.
In other words, $\gamma$ is proper if there is an closed
submanifold $W_{\gamma}^\st(A)$ such that
\begin{enumerate}
\item
$\partial W_{\gamma}^\st(A)=\gamma$, and
\item
$f_0(W_{\gamma}^\st(A))\subset int(W_{\gamma}^\st(A))$.
\end{enumerate}

We refer to the closed set $W_{\gamma}^\st(A)$ as the stable manifold
\textit{starting at }$\gamma$.

Similarly, for the unstable manifold, a submanifold $\sigma\subset W^\un(A)$
is proper it is proper for $f_0^{-1}$. However, in this case we define the
unstable manifold \textit{up to} $\sigma$, denoted by $W_{\sigma}^\un(A)$,
as the \emph{interior} of the local manifold that corresponds to $f_0^{-1}$.
\end{defi}

Notice that the definition is not symmetric, because $W_{\gamma}^\st(A)$ is a
closed subset of $W^\st(A)$, while $W_{\sigma}^\un(A)$ is open in $W^\un(A)$.
The asymmetry is just a technicality in order to simplify some proofs.

\begin{defi}[Fundamental Domain]
Let $A$ be a hyperbolic invariant set for $f_0$.
A submanifold with boundary $P$ is a fundamental domain of
$W^\st(A)$ if there exists some proper boundary $\gamma\in W^\st(A)$
such that
\[
P=W_{\gamma}^\st(A)\setminus W_{f_0\circ\gamma}^\st(A)\;.
\]
Equivalently, a fundamental domain in $W^\un(A)$ is a manifold with boundary
of the form
\[
P=W_{\sigma}^\un(A)\setminus W_{f_0^{-1}\circ\sigma}^\un(A),
\]
where $\sigma$ is a proper boundary in $W^\un(A)$.
\end{defi}

In each case, the fundamental domain is a manifold with boundary
$\partial\mathcal{P=}\gamma\cup f_0(\gamma)$. An immediate consequence of
the definition is that all the forward and backward iterations of a
fundamental domain are also fundamental. It is easy to see that proper
boundaries always exist, and in fact, the unstable manifold can be decomposed
as the disjoint union of fundamental domains:
\[
W^\st(A) \setminus A = \bigcup_{k\in \Zset }f_0^{k} (P)\;.
\]

The importance of fundamental domains is that much of the information about
the entire manifold can be found by looking only at these submanifolds.
For example, as discussed in \cite{LomeliM03},
the topology of the intersections of $W^\un$ and $W^\st$ can be studied by
restricting to $P$.

\begin{pro}\label{pro:sony}
Let $M$ be a simply connected manifold and
$f_\epsilon:M \to M$ a family of volume-preserving maps such that
$f_0$ has a saddle connection $\Sigma$ with fundamental domain $P$.
If $\omega_\eta$ is the $(n-1)$-form defined in~\eref{eq:omegaEta} and
$M_\eta$ is the Melnikov function~\eref{eq:melnikovVolume},
then $\int_{P}M_\eta\omega_\eta=0$.
\end{pro}

\proof
The fundamental domain $P$ is a submanifold with boundary,
such that $\partial P=\gamma\cup f_0(\gamma)$,
where $\gamma$ is a closed curve that does not intersect $f_0(\gamma)$.
If we give an orientation $\left[P\right]$ to $P$,
the induced orientation on the boundary satisfies
$\left[\gamma\right] = -\left[f_0(\gamma)\right]$.
According to proposition~\ref{pro:exact}, the form $\Interior{\mathcal{F}_0} \Omega$
is exact.
Thus there exists an $(n-2)$-form $\Xi$ such that
${\rm d} \Xi = \Interior{\mathcal{F}_0} \Omega$.
 From~\eref{eq:MelnikovSums} we conclude that
\[
\Interior{\mathcal{D}} \Omega =
\Interior{ \sum_{k\in\Zset} (f_0^\ast)^k \mathcal{F}_0 } \Omega =
\sum_{k\in\Zset} (f_0^\ast)^k \Big(\Interior{\mathcal{F}_0} \Omega\Big) =
{\rm d} \sum_{k\in\Zset} (f_0^\ast)^k \Xi =
{\rm d} \Psi
\]
where $\Psi = \sum_{k\in\Zset} (f_0^\ast)^k \Xi  $. Finally, from
definition~\ref{def:java}, the Stokes's theorem, and the invariance
$f_0^\ast \Psi =\Psi$, we get that the integral
\[
\int_{P} M_\eta \omega_\eta = \int_{P}
\Interior{\mathcal{D}} \Omega = \int_{P}{\rm d} \Psi =
\int_{\partial P} \Psi = \int_{\gamma}\Psi -
\int_{\gamma} f_0^\ast \Psi
\]
vanishes.
\qed

The previous result implies that the stable and unstable manifolds
of a perturbed saddle connection necessarily intersect. Examples
of such intersections were computed for the case $M =
\Rset^3$---where the hypothesis of proposition~\ref{pro:sony} are
satisfied---in \cite{LomeliM00a,LomeliM03}.

\section{Conclusion and future research}

We have studied a general theory of the Melnikov method that can be applied to
many different settings.
The formula~\eref{eq:MelnikovSums} for the Melnikov displacement $\mathcal{D}$
generalizes many of the classical methods that use a normal vector field
to measure the displacement with respect to a natural direction.
For example,
when the saddle connection is defined as the level set of a first integral $I$,
the classical Poincar\'{e} function $M_I = {\rm d} I(\mathcal{D})$
measures the splitting as the rate of change of the first integral.
If there is a Riemannian structure and an associated adapted normal vector
field $\eta$,
then the classical Melnikov function $M_\eta = \langle \eta, \mathcal{D} \rangle$
measures the rate of change of the splitting in this normal direction.
For the case of exact symplectic maps with saddle connections between
hyperbolic fixed points,
the Melnikov potential $L$ is defined on a Lagrangian submanifold and
acts as a generator for the displacement:
${\rm d} L = \Interior{\mathcal{D}} \omega$.

This Melnikov theory can be extended to other situations
and can be applied in many problems.
For instance, one can study billiard dynamics inside a perturbed
ellipsoid, following a program initiated in~\cite{DelshamsFR01,BolotinDR04}.
It turns out that the billiard map inside an ellipsoid is an exact symplectic
diffeomorphism defined on the cotangent bundle of the ellipsoid,
which has a two-dimensional normally hyperbolic invariant manifold
with a three-dimensional saddle connection.
Therefore,
ellipsoidal billiards represent a strong motivation for a more
detailed study of general normally hyperbolic invariant manifolds
in a symplectic framework.
This is a work in progress.

In discrete volume geometry there are many open Melnikov problems,
since the application of Melnikov methods to volume-preserving maps
began just a few years ago~\cite{LomeliM00a,LomeliM03}.
We plan to continue this program in several ways.
As a first step, we plan to obtain bounds on the number of primary
heteroclinic orbits in terms of the degree of the polynomial perturbation~\cite{LomeliR07}.

\ack HL was supported in part by Asociaci\'{o}n Mexicana de Cultura.
JDM was supported in part by NSF grant DMS-0202032 and by the
Mathematical Sciences Research Institute in Berkeley. RR-R was
supported in part by MCyT-FEDER grant MTM2006-00478. Useful
conversations with Guillermo Pastor, Amadeu Delshams, Richard
Montgomery, Clark Robinson, Renato Calleja and Rafael de la Llave
are gratefully acknowledged.

\end{document}